\documentclass[aps,pra,twocolumn,showpacs,floatfix,superscriptaddress,10pt]{revtex4-1}

\usepackage{graphicx}
\usepackage{amsfonts}
\usepackage{amssymb}
\usepackage{amsmath}
\usepackage{color,verbatim}

\begin{document}

\title{Dragging A Defect in a Droplet Bose-Einstein Condensate}

\author{S. Saqlain}
\affiliation{Department of Mathematics and Statistics,
University of Massachusetts, Amherst MA 01003-4515, USA}

\author{Thudiyangal Mithun}
\affiliation{Department of Mathematics and Statistics,
University of Massachusetts, Amherst MA 01003-4515, USA}

\author{R. Carretero-Gonz{\'a}lez}
\affiliation{Nonlinear Dynamical Systems
Group,\thanks{\texttt{URL}: http://nlds.sdsu.edu}
Computational Sciences Research Center, and
Department of Mathematics and Statistics,
San Diego State University, San Diego, California 92182-7720, USA}

\author{P.G. Kevrekidis}
\affiliation{Department of Mathematics and Statistics,
University of Massachusetts, Amherst MA 01003-4515, USA}

\begin{abstract}
In the present work we consider models of quantum droplets in the presence of a defect
in the form of a laser beam moving through the respective condensates
including the Lee-Huang-Yang correction. Our analysis 
features separately an exploration of the existence, stability, bifurcations and dynamics
in 1D, 2D and 3D settings. In the absence of an analytical solution of the problem, we 
provide an analysis of the speed of sound and observe how the states traveling with the
defect may feature a saddle-center bifurcation as the speed or the strength of the defect
is modified. Relevant bifurcation diagrams are constructed systematically,
and the unstable states, as well as the dynamics past the existence of stable states
is monitored. The connection of the resulting states with dark solitonic patterns in 1D,
vortical states in 2D and vortex rings in 3D is accordingly elucidated.
\end{abstract}

\maketitle

\section{Introduction}

The study of atomic condensates has been a focal point at the interface of
atomic, nonlinear optical and wave physics during the last three decades
since the experimental realization of Bose-Einstein condensates (BECs) 
of dilute alkali gases~\cite{pethick2008bose,stringari}.
From the nonlinear perspective, a wide array of studies
has taken place in this context, ranging from the exploration
of one-dimensional (1D) nonlinear waves in the form of dark solitons 
to two-dimensional (2D) vortices and their lattices and 
finally in the three-dimensional (3D) setting 
and the examination of vortex lines and rings~\cite{siambook}. 

A particular topic that has attracted considerable attention over the
years, including in a wide range of experiments
has been the dragging of a defect (in the form
of a light beam) through the condensate
and the observation of the ensuing dynamics, 
especially so if the defect moves with a speed
larger than the local speed of sound (violating
the well-known Landau criterion)~\cite{PhysRevLett.83.2502,PhysRevLett.85.2228,PhysRevLett.99.160405} and accordingly producing nonlinear excitations.
Indeed, the relevant subject of both the
associated instability and the resulting 
pattern formation has been of intense
interest since early on~\cite{PhysRevLett.69.1644,PhysRevLett.82.5186,hakim} and continues to lead to variations on the
relevant theme~\cite{PhysRevA.64.033602,PhysRevA.66.013610}, e.g., involving trapping~\cite{PhysRevA.68.043614,PhysRevA.92.013627}, 
for oscillating obstacles~\cite{PhysRevA.70.013602},
in different dimensions~\cite{CARRETEROGONZALEZ2007361}, in periodic rings~\cite{Syafwan_2016},
for a larger number of components~\cite{PhysRevA.75.055601,PhysRevA.79.043603}, or in the setting of polariton condensates
involving dissipation and pumping processes~\cite{PhysRevB.86.165304} among many
others, including recently beyond mean-field 
effects~\cite{PhysRevA.98.013632}.

On the other hand, a direction that has
gained considerable traction lately 
has to do with the emergence of an effective
new type of matter wave in the form of
quantum 
droplets~\cite{petrov2015quantum,Petrov_2016}.
The relevant physical setting involves
two-component (binary) BECs in which the
(effectively nonlinear) interaction 
interplay involves intra-component 
repulsion and inter-component attraction
(which slightly exceeds the former).
It is in this system that the 
well-established Lee-Huang-Yang (LHY) 
quantum correction~\cite{lee1957eigenvalues}
can be used to incorporate the
averaged beyond-mean-field effect
of quantum fluctuations in the 
dynamical description, while 
competing with the mean-field effects. 
A key role of the relevant correction
is that of preventing the potential BEC collapse of the mean-field realm
in higher dimensions. 
Such beyond mean-field fluctuations
have been found to be attractive in
1D settings, while they contribute in a
repulsive way beyond 1D.

A key appeal of these predictions 
and the associated study of quantum
droplets is that they have led to a number of experimental implementations of
this system{~\cite{cabrera2018quantum,cheiney2018bright,Semeghini2018self,Ferioli_2019,fort}, 
while such droplets were originally realized in dipolar settings~\cite{Chomaz2016dipolar,Ferrier2016dipolar}.} 
Indeed, not only have individual droplets
been observed but also their interactions
in the form of collisions (both resulting
in slow-collision mergers and in faster
quasi-elastic events) have been reported in~$^{39}$K ~\cite{Ferioli_2019}. Beyond homonuclear
settings, heteronuclear droplets of 
$^{87}$Rb and $^{41}$K have also been shown to be 
long-lived~\cite{fort}. 
These substantial experimental
findings have, in turn, motivated 
various theoretical studies, such as, e.g.,
the one involving vortical droplet
patterns~\cite{yongyao}, the one of 
semi-discrete ones with  or without vorticity~\cite{semidiscrete}, their
dynamics in optical lattices~\cite{Morera2020QD},
their modulational stability~\cite{mithun2019inter}
or the
case of 3D such structures in Ref.~\cite{kartashov}.
A relatively recent recap of theoretical
and experimental activity in this field can be
found in Ref.~\cite{luo2020new}.

Our aim in the present work is to combine 
the above two central directions, namely the
study of the potential dragging in the form
of a defect and the exploration of models
of quantum droplets. An interesting feature
of the latter models is their distinct,
yet well-established form in each of
the three dimensions (1D, 2D and 3D)~\cite{luo2020new}. 
Here, we explore, in the case of symmetric components,
each one of these settings separately,
developing an analysis of the corresponding
equilibrium states and obtaining the
respective sound speeds as a function of the
chemical potential. Subsequently, in the
spirit of the work of Ref.~\cite{hakim}, we
consider a Gaussian defect (as a finite
width emulation of a $\delta$-function one).
We then use a systematic bifurcation analysis
to explore the stable and unstable configurations
pertaining to the moving defect, when ones such
exist (below the local speed of sound) as a function
of the dragging speed and the defect strength.
While in the case of the cubic nonlinear
Schr{\"o}dinger (NLS) model, the existence of an
explicit dark soliton solution enables an
analytical calculation of the relevant 
bifurcation curve, here the absence of
such an analytical coherent structure expression
leads us to identify the relevant curves 
numerically. We construct the corresponding
two-parameter diagram and connect it with
the speed of sound (to which  the critical
speed tends, as the height of the defect
goes to 0). We perform such calculations
both for 1D and also for higher dimensions
(2D and 3D). In the latter, the systematic
bifurcation curves present interesting 
features including an unstable branch 
bearing vortical states moving along with
the defect. Interestingly, such states exist
even for cubic nonlinearities. Finally, in
3D, the vortex pairs are replaced by vortex
ring structures which are moving with the
defect, a pattern of interest in its own right.

Our presentation will be structured as follows.
In Sec.~\ref{sec:theory}, we will present the models
at hand based on symmetric populations between the
two components of the binary mixture. 
We will also derive the speed of sound corresponding to
the different dimensionalities.
In Sec.~\ref{sec:numerics} we will present,
for the 1D, 2D and 3D settings,
the corresponding equilibria and 
subsequently explore the bifurcation 
structure of stable and unstable states,
the saddle-center bifurcation that they
feature and the corresponding dynamics
both for parameters before and for ones
after the bifurcation. 
Finally, Sec.~\ref{sec:conclu} will contain a summary
of our findings and a corresponding list
of possible directions for future work.
In the Appendix, we revisit similar features
for the cubic nonlinearity (higher-dimensional)
case for completeness.

%

\section{Model equations and theoretical analysis}
\label{sec:theory}

In the analysis that follows we focus our attention on the simpler, so-called
symmetric, case where there is no population imbalance between 
the two components of the BEC binary mixture. 
Different aspects of asymmetric population mixtures in 1D, 2D, 
and 3D have been considered, respectively, in 
Refs.~\cite{mithun2019inter,yongyao,kartashov}.
In 3D and under the assumption of symmetric populations, both BEC
components are identical and can be described by a single wavefunction 
$\psi(\vec{r},t)$ satisfying the following dimensionless
Gross-Pitaevskii (GP) equation:
\begin{equation}
\label{eq:GP3D}
i\partial_t \psi = -\frac{1}{2}\nabla^2 \psi 
+ {\cal{N}}(\psi)
- \mu \psi + {\cal{V}}(\vec{r},t)\psi,
\end{equation}
where $\mu$ is the chemical potential, ${\cal{V}}(\vec{r},t)$ is the
external potential, and ${\cal{N}}(\psi)$ is the effective nonlinearity. 
Note that in the symmetric case under consideration, 
$\mu$, ${\cal{V}}(\vec{r},t)$, and ${\cal{N}}(\psi)$ are equal for 
both binary components.
The crucial aspect when considering the LHY correction is that the 
effective nonlinearity deviates from the usual cubic one given by
$|\psi|^2\psi$ and that it takes different forms depending on the effective
dimensionality of the system~\cite{luo2020new}.
In particular, for the different effective spatial dimensions the
nonlinearity takes the form:
\begin{equation}
{\cal{N}}(\psi)=
\left\{
\begin{array}{ll}
|\psi|^2 \psi - |\psi| \psi & {\rm in~1D},
\\[1.0ex]
\ln{\left(|\psi|^2\right)} |\psi|^2 \psi & {\rm in~2D},
\\[1.0ex]
g_1 |\psi|^2\psi + |\psi|^3\psi & {\rm in~3D},
\end{array}
\right.
\end{equation}
where $g_1$ can be positive or negative.
These three cases will be considered herein.

We consider an impurity of fixed shape $V$ moving across the BEC 
at velocity $c$ along, without loss of generality, the $x$-direction 
such that ${\cal{V}}(\vec{r},t)=V(x-ct,y,z)$.
Thus, to be able to track steady states arising from the inclusion
of the impurity, we cast the evolution equations in a co-moving
reference frame $\xi=x-ct$ where the impurity is stationary, yielding.
\begin{equation}
\label{eq:GP3DcoM}
i\partial_t A -ic\,\partial_xA= -\frac{1}{2}\nabla^2 A 
+  {\cal{N}}(A)
- \mu A + V(\vec{r})A,
\end{equation}
where we relabeled $\xi \rightarrow x$ and consider 
$A(x,y,z,t)=\psi(x-ct,y,z,t)$.
We now study the different cases corresponding to 1D, 2D, and 3D.

\subsection{One-dimensional setting}
\label{sec:theory:1D}

In practice, the BEC needs to be formed in the presence of a confining potential (in addition
to the potential describing the running impurity). Provided this confining
potential has very strong confinements in two directions, let us say
along $y$ and $z$, the dynamics of the BEC can be well approximated by 
the 1D version of the GP equation~(\ref{eq:GP3DcoM})~\cite{luo2020new}.
In this quasi-1D case, where the transverse $y$ and $z$ directions have
been factored (or better said, averaged) out,  
the BEC wavefunction $A(x)$, in the co-moving reference frame,
satisfies the (effective) 1D equation
\begin{equation}
\label{eq:1}
i\partial_t A -ic\,\partial_xA
 = -\frac{1}{2}\partial_{xx} A + |A|^2A - |A|A - \mu A + V(x)A,
\end{equation}
where $V(x)$ now represents the 1D stationary profile of the running
impurity in the co-moving reference frame.

In the homogeneous case, 
when the defect is absent, i.e., $V(x)=0$, Eq.~(\ref{eq:GP3D}) admits a 
homogeneous, space-independent, stationary steady state 
$|A(\vec{r},t)|=|\alpha|$ such that
\begin{eqnarray}
\label{eq:A1D}
|\alpha|^2 -|\alpha| - \mu = 0
\quad\Rightarrow\quad
|\alpha| = \frac{1\pm \sqrt{1+4\mu}}{2},
\end{eqnarray}
where both sign solutions exist for $\mu>-1/4$ and the one with the
$(-)$ sign also for $\mu<0$.
Looking now for co-moving steady states for non-zero defects of 
Eq.~(\ref{eq:1}) of the general form 
\begin{eqnarray}
\label{eq:ARphi}
A(x) = R(x)\,e^{i\phi(x)},
\end{eqnarray}
where $R$ and $\phi$ are real functions, yields
\begin{eqnarray}
 cR_x &=& \frac{1}{2} (2R_x\phi_x+R\phi_{xx}),
 \label{eq:2}
\\[1.0ex]
\notag
 cR\phi_x&=&-\frac{1}{2} (R_{xx}-R\phi_x^2)+R^3-R^2-\mu R+VR.
\end{eqnarray}
Integrating the first equation yields
$$
\phi_x=c\left(1-\frac{|\alpha|^2}{R^2}\right),
$$
which, after substituting in the second equation yields,
\begin{equation}
\label{eq:5}
 R_{xx} = c^2\left(-R+\frac{|\alpha|^2}{R^3}\right) +2R^3-2R^2-2\mu R+2VR.
\end{equation}

Since we assume that the defect is localized $V(x \rightarrow \pm\infty) \rightarrow 0$, we must 
require that $R(x) \rightarrow |\alpha|$ as $x \rightarrow \pm \infty$. 
Then, linearizing $R(x)$ as  $R(x) = |\alpha| + r(x)$ for $x$ away 
from the center, yields
\begin{equation}
 r_{xx}=2r(|\alpha|+2\mu-2c^2),
\end{equation}
implying that the speed of sound $c_s$ for the 1D setting
is given by 
\begin{equation}
\label{eq:cs1D}
c_s=\sqrt{\frac{|\alpha| + 2\mu}{2}}.
\end{equation}
in analogy with the cubic nonlinearity calculation of~\cite{hakim}.
 {It is to be noted that this relation, Eq.~(\ref{eq:cs1D}), can also be obtained from the pressure $p = n^2 \partial_n (E_{0v}/n)$, where the ground energy per volume is $E_{0v} = \frac{1}{2} n^2 - \frac{2}{3} n^{3/2}$ and $n=|\alpha|^2$ is the density~\cite{pethick2008bose}. Additionally, as per the Landau criterion~\cite{pethick2008bose,stringari}, this represents the critical velocity required to create an excitation by a moving obstacle with velocity $c$ through a uniform system.}

\subsection{Two-dimensional setting}
\label{sec:theory:2D}

Let us now consider the 2D setting. In this case, assuming a
strong confinement along, let us say, the $z$ direction and by appropriately
averaging across this transverse direction, the 2D wavefunction 
in the co-moving reference frame evolves according to~\cite{luo2020new}
\begin{multline}
\label{eq:6}
i\partial_t A -ic\,\partial_xA
= -\frac{1}{2}(\partial_{xx} + \partial_{yy})A \\
+ \ln{\left(|A|^2\right)} |A|^2 A- \mu A + V(x,y)A.
\end{multline}
Note, importantly, that the nonlinearity, modified by the LHY 
correction term, assumes a somewhat unusual logarithmic form 
in the 2D setting.
In this case, the homogeneous steady state $|A(\vec{r},t)|=|\alpha|$
solves the transcendental equation
\begin{eqnarray}
\label{eq:A2D}
|\alpha|^2 \ln{|\alpha|^2} - \mu = 0.
\end{eqnarray}
A similar analysis to the one done in 1D, see Appendix~\ref{app:2D}, 
yields the speed of sound for the 2D setting as
\begin{equation}
\label{eq:cs2D}
c_s = \sqrt{|\alpha|^2+\mu}.
\end{equation}
%

\subsection{Three-dimensional setting}
\label{sec:theory:3D}

Finally, let us now consider the 3D setting. In this case, we use
directly Eq.~(\ref{eq:GP3D}) in the co-moving reference fame:
\begin{eqnarray}
i\partial_t A -ic\,\partial_xA &=& -\frac{1}{2}\nabla^2 A  + g_1 |A|^2 A 
\notag
\\[1.0ex]
\label{eq:12}
&& + |A|^3A- \mu A + V\,A,
\end{eqnarray}
where we have allowed the intrinsic cubic nonlinear term to
be tuned from attractive $g_1<0$ to repulsive $g_1>0$.
The homogeneous steady state $|A(\vec{r},t)|=|\alpha|$ now yields 
a cubic equation for the amplitude:
\begin{eqnarray}
\label{eq:A3D}
|\alpha|^3 + g_1 |\alpha|^2 - \mu = 0.
\end{eqnarray}
A similar analysis to the one done in 1D, see Appendix~\ref{app:3D}, 
yields the expression for the speed of sound in the 3D setting
\begin{equation}
\label{eq:cs3D}
c_s = \sqrt{\frac{|\alpha|^3+2\mu}{2}}.
\end{equation}
It is interesting to note that the expression for the speed of sound in
Eq.~(\ref{eq:cs3D}) seems not to depend on the sign of $g_1$. Nonetheless,
it is important to note that, although indeed Eq.~(\ref{eq:cs3D}) does not 
explicitly depend on $g_1$, it does so through the (dependence on $g_1$ of the) background level 
$|\alpha|$ as per Eq.~(\ref{eq:A3D}).

\section{Numerical results}
\label{sec:numerics}

In this section we corroborate the predictions for the speed
of sound of Sec.~\ref{sec:theory} and follow the different
steady states and their dynamics for the
different dimensionalities.
For the numerical results, we use a standard finite difference 
discretization of second order in space and 4th order Runge-Kutta stepping
in time.
1D, 2D, and 3D results are typically obtained, respectively, for domains 
$x\in[-200,200]$, $(x,y)\in[-30,30]^2$, and $(x,y,z)\in[-30,30]^3$,
with corresponding spatial discretizations such that
lower $dx$ was checked not to provide notable differences,
and a time step $dt$
below the stability threshold given in Ref.~\cite{Caplan201324}.
The steady states are obtained by standard fixed point iteration methods
and the solution branches as parameters are varied were obtained using
pseudo-arclength continuation~\cite{doedel2007lecture}.
The bifurcation analysis presented has leveraged the use of the 
\texttt{Julia}'s bifurcation package \texttt{BifurcationKit}~\cite{veltz2020},
especially so for our 1D and 2D results. 
All the steady state and dynamics shown here are depicted in the
co-moving reference where the defect is stationary.

\subsection{One-dimensional setting}
\label{sec:num:1D}

For the 1D setting governed by Eq.~(\ref{eq:1}), we consider a narrow,
Gaussian laser beam defect that runs through the condensate at velocity $c$.
In the co-moving reference frame, this 1D defect takes the form
$$
V_{\rm 1D}(x) = \frac{\lambda}{\sqrt{2\pi \epsilon_x^2}} \exp \left(\frac{-x^2}{2\epsilon_x^2}\right),
$$
where $\lambda$ is the strength (intensity) of the defect and $\epsilon_x$
characterizes its waist (width). For our numerics, corresponding to the adimensionalized
model of Eq.~(\ref{eq:1}), we choose a chemical potential of $\mu=1$ and
a defect with waist $\epsilon_x=0.2$.

\begin{figure}[ht!]
\centering
\includegraphics[width=\columnwidth]{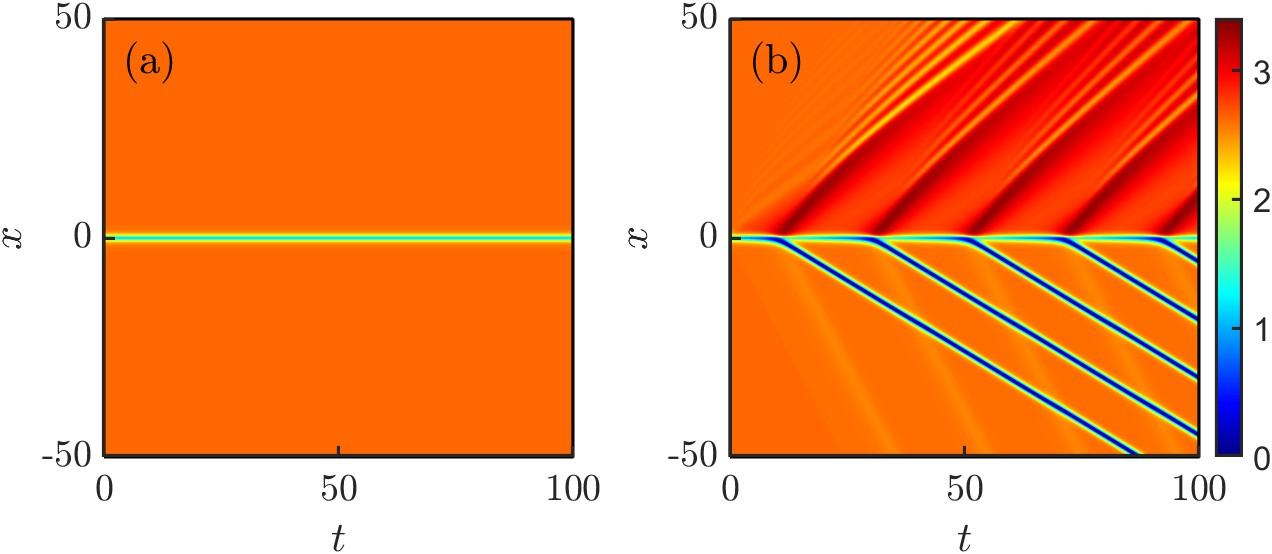}
\caption{Dynamics ensuing from a 1D defect running through the BEC
for $\mu=1$ and $\lambda=0.75$.
(a) Subcritical case corresponding to a velocity $c=0.647$ below the 
speed of sound.  The resulting steady state is stable.
(b) Supercritical case corresponding to a velocity $c=0.7$ above the 
speed of sound. The resulting dynamics cannot feature a stationary
state and, consequently, gives rise 
to a periodic emission of dark solitons in its wake.}
\label{fig:evol_1d}
\end{figure}

For large enough $c$, namely past the speed-of-sound threshold, as it
is the  case for the pure cubic NLS nonlinearity  without the LHY 
correction~\cite{hakim,CARRETEROGONZALEZ2007361}, 
the co-moving steady state can no longer be supported (it has
terminated in a saddle-center bifurcation as we show below) 
and consequently emits a periodic
train of dark solitons in its wake. Essentially, the emission of
the solitary waves renders the dynamics temporarily and locally
subcritical. Yet, once the solitary wave has moved enough upstream,
the defect region becomes supercritical anew and yields an additional
dark solitary structure, eventually resulting in the production of
a train thereof.
A example of this behavior is depicted in Fig.~\ref{fig:evol_1d}.
The left panel shows how a defect running at low enough velocities
gives rise to a stable co-moving steady state. However, past a critical
value of the speed, as shown in the right panel, this steady state 
disappears through a bifurcation as illustrated below, resulting in the
production of a periodic train of dark solitons.
As the speed of the defect is increased further 
the
spacing between the dark soliton emissions decreases.

\begin{figure}[ht!]
\centering
\baselineskip 0pt
\includegraphics[width=\columnwidth]{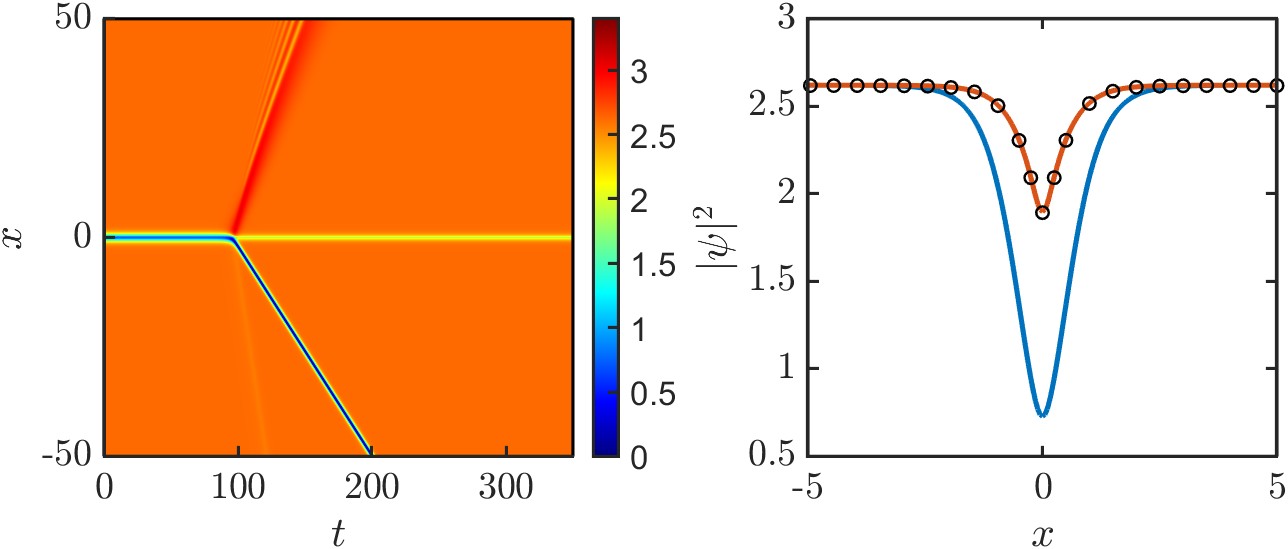}
\caption{Evolution of the 1D unstable solution for a defect velocity 
$c=0.7$ below the speed of sound for $\mu=1$ and $\lambda=0.4$.
The left panel shows the evolution where the deeper unstable steady
state solution sheds a single dark soliton and settles to an apparently
stable shallower solution.
The right panel confirms that the initial unstable steady state (blue curve)
decays (after the emission of the single dark soliton) to a shallower
solution (red curve) that precisely corresponds to the stable steady
state solution for this speed (circles).
}
\label{fig:evol_1d_upper}
\end{figure}

\begin{figure}[ht!]
\centering
\includegraphics[width=0.8\columnwidth]{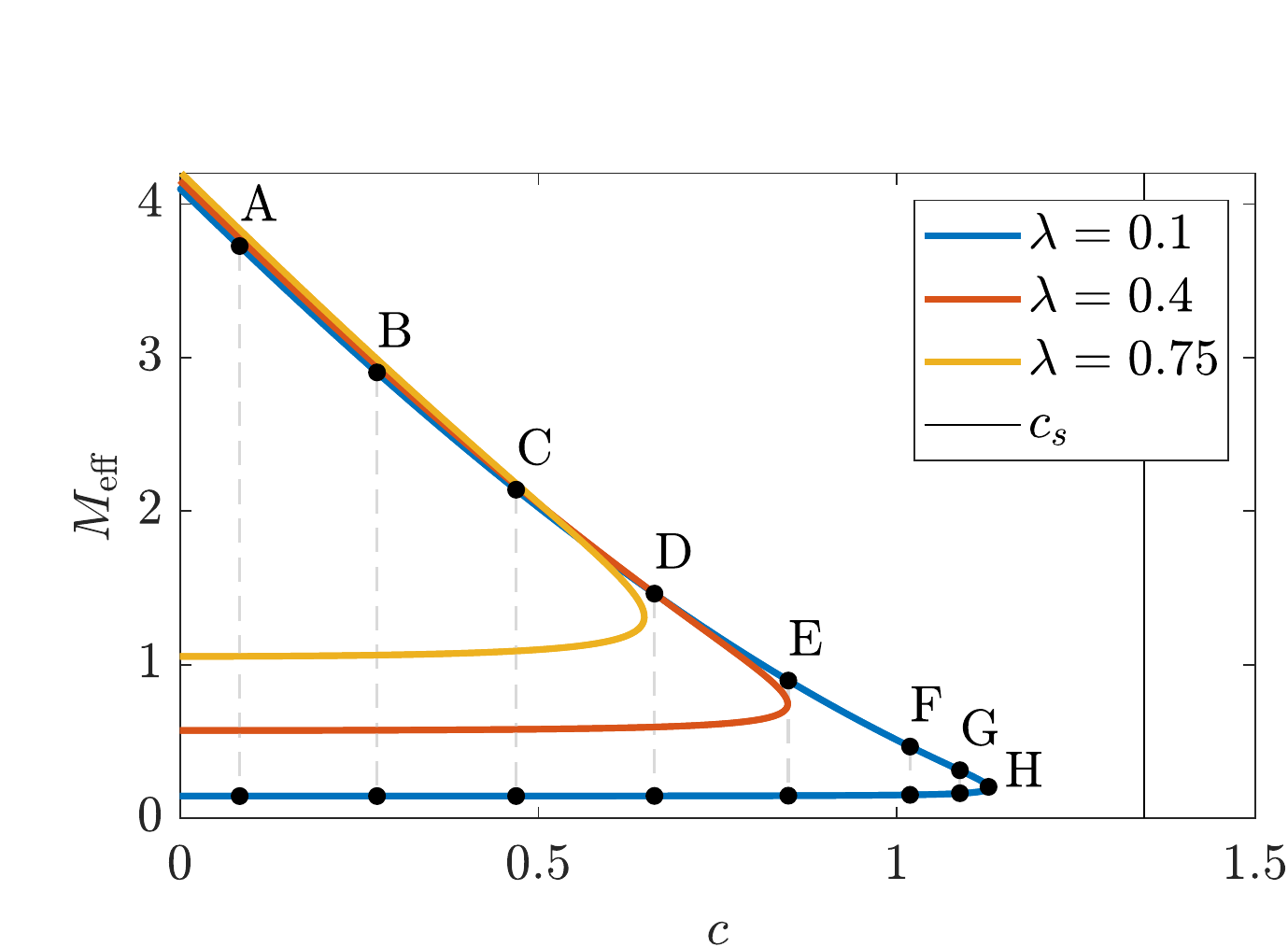}
\\[3.0ex]
\includegraphics[width=\columnwidth]{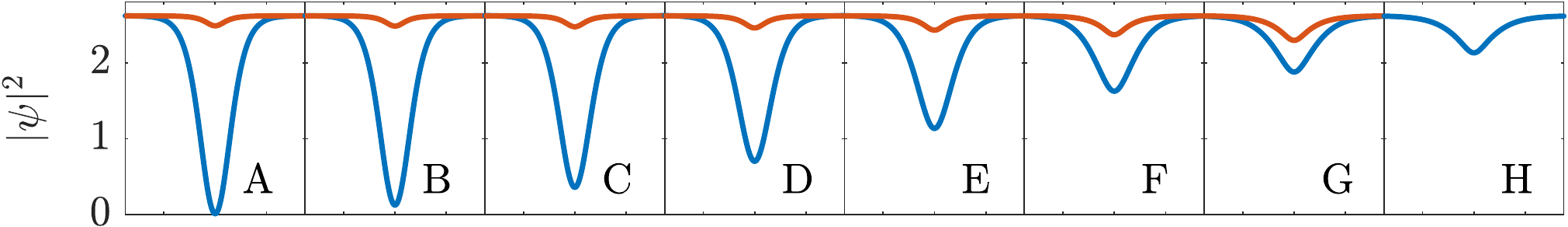}
\\[2.0ex]
\includegraphics[width=\columnwidth]{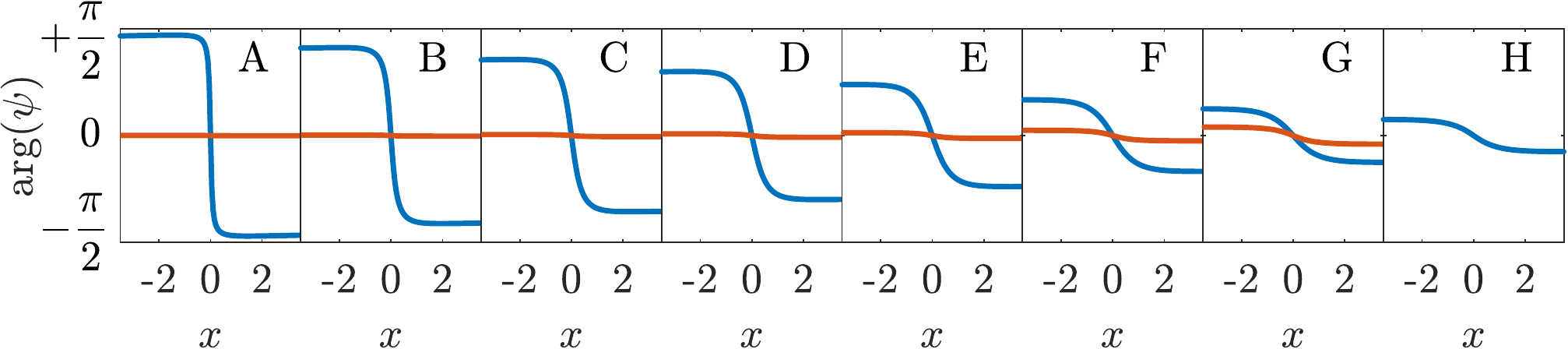}
\caption{(Color online)
Top:
Bifurcation diagram for the stable (lower part of the
corresponding branches) and unstable (upper part of the branches)
branches in 1D as the velocity of the impurity $c$ is varied for $\mu=1$
and for three values of the impurity strength $\lambda$ as indicated 
in the legend.
The effective mass of the solution (see text) is plotted vs.~$c$.
As $c$ reaches a threshold value, the upper and lower branches
coalesce in a saddle-center bifurcation (see point H).
For values of $c$ larger than this threshold, 
there no longer exists a stable stationary state and hence the 
time evolution dynamics periodically sheds dark
solitons in its wake (see right panel in Fig.~\ref{fig:evol_1d}).
The vertical black line corresponds to the theoretical prediction of
speed of sound in Eq.~(\ref{eq:cs1D}).
Middle (bottom):
Corresponding density (phase) profiles at the different values of $c$ indicated by the 
black dots in the top panel. Each subpanel depicts the corresponding 
unstable (deep; see blue curve) and stable (shallow; see red curve) solutions.
}
\label{fig:bif_1d}
\end{figure}

In order to numerically determine the critical speed for dark soliton emission,
we study the steady states that exist as the defect velocity is varied. 
In particular, as it is the case for the pure NLS case without the LHY correction (see for instance 
Ref.~\cite{hakim}), 
for values of $c$ below the speed-of-sound threshold, there
exist two steady state solutions: a relatively shallow stable
state and a relatively deep unstable steady state.
For example, Fig.~\ref{fig:evol_1d_upper} depicts the evolution of the
unstable solution for $c$ below the speed of sound. This deeper solution
decays, after the emission of a single dark soliton, to a shallower
solution that precisely corresponds to the stable solution.
We can now trace the families of unstable (deeper) and stable
(shallower) solutions as the parameters of the system are varied.
In particular, in Fig.~\ref{fig:bif_1d} we follow using pseudo-arclength
continuation in $c$ these solution branches for several values of the
defect strength $\lambda$. To monitor the solutions,
we use the effective mass of the solution:
\begin{equation}
\label{eq:Mass}
M_{\rm eff} = \int_{-\infty}^{+\infty} 
\left[|\alpha|^2-|\psi|^2\right]\, dx,
\end{equation}
where $|\alpha|$ is the background level as defined in Eq.~(\ref{eq:A1D}).
According to this definition, the deeper the solution the higher the effective mass.

\begin{figure}[ht!]
\centering
\includegraphics[width=0.9\columnwidth]{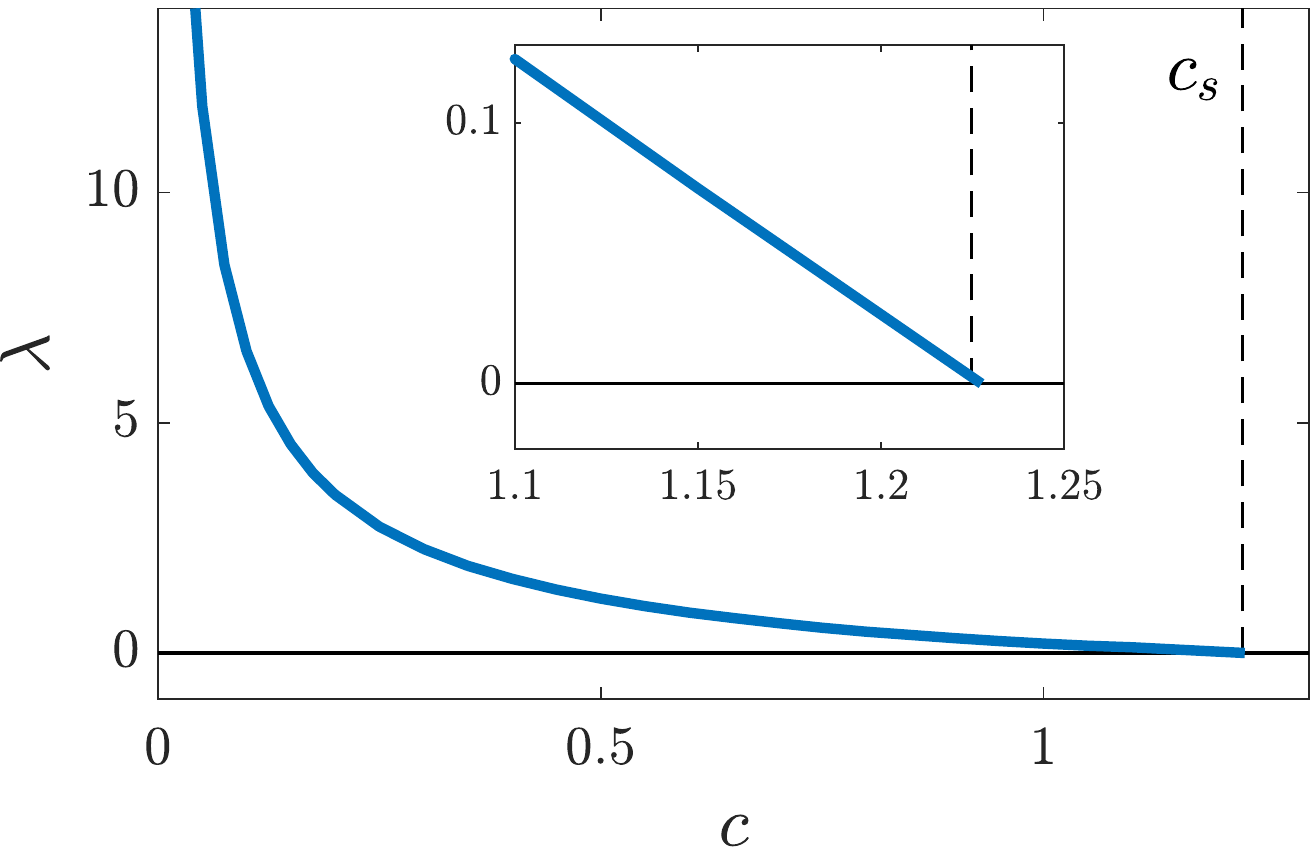}
\caption{Critical values of defect strength $\lambda$ vs.~defect speed $c$ 
for the 1D case.
For each value of $c$ we extract the value of $\lambda$ where the saddle-center
bifurcation occurs (collision between the deep and shallow steady state solutions).
The vertical dashed line corresponds to the theoretical prediction of the
speed of sound in Eq.~(\ref{eq:A1D}).
The inset shows a zoomed-in version for small defect strengths.
}
\label{fig:lamc_1d}
\end{figure}

As can be seen in Fig.~\ref{fig:bif_1d}, the deep and shallow solutions,
corresponding respectively to the upper and lower solution branches, coalesce 
(see the turning point H) as the defect velocity $c$ is increased. At a critical value of 
$c$, a saddle-center bifurcation ensues where the two solutions collide.
Past this critical value of $c$, 
the system bears no stable solution and, hence, as explained above,
periodic emission of dark solitons takes place
at the wake of the impurity as seen
in the right panel in Fig.~\ref{fig:evol_1d}.
In Fig.~\ref{fig:bif_1d} we also depict the theoretical prediction for the
speed of sound of Eq.~(\ref{eq:cs1D}) (see vertical black line). 
As seen in Fig.~\ref{fig:bif_1d}, the defects always have a critical speed
that is below the speed of sound. This suggests that defects will
emit dark soliton trains for values slightly below the speed of sound.
Nonetheless, the figure also suggests that the critical speed value 
tends to approach the speed of sound as the defect strength $\lambda$ decreases.
In fact, as the computation of the speed of sound in Sec.~\ref{sec:theory}
relies on small perturbations, the results will be valid in the
limit of $\lambda\rightarrow 0$.
To elucidate this connection, we depict in Fig.~\ref{fig:lamc_1d}
the values of the defect strength $\lambda$ where the saddle-center 
collision between the deep and shallow steady state occurs.
As it can be corroborated (see inset), the critical velocity tends to 
coincide with the theoretical speed of sounds $c_s$ of Eq.~(\ref{eq:A1D})
when the defect strength $\lambda$ tends to zero.

\subsection{Two-dimensional setting}
\label{sec:num:2D}

We now proceed in a similar way as for the 1D case of the previous section
but now for the 2D model of Eq.~(\ref{eq:6}). Here we choose a 2D defect
in the form of a `bar' that is thin in the direction of the defect movement
(namely the $x$-direction) and relatively wide in the transverse direction.
Specifically, the 2D defect is taken to be
$$
V_{\rm 2D}(x,y) = 
V_{\rm 1D}(x)\,
\exp \left(\frac{-y^2}{2\epsilon_y^2}\right).
$$
Note that we do not use the typical normalization constant for a 2D Gaussian
as we are considering the thin defect $x$-direction the one that drives the
nucleation of dark solitons and thus, to best compare with the 1D results,
we use the same normalization prefactor as for the 1D case.
For our numerical results below, we choose a thin defect bar with 
$\epsilon_x=1/\sqrt{2}\approx 0.707$ and a relatively wide lateral extent
with $\epsilon_y=5$.

\begin{figure}[ht!]
\centering
\includegraphics[width=0.85\columnwidth]{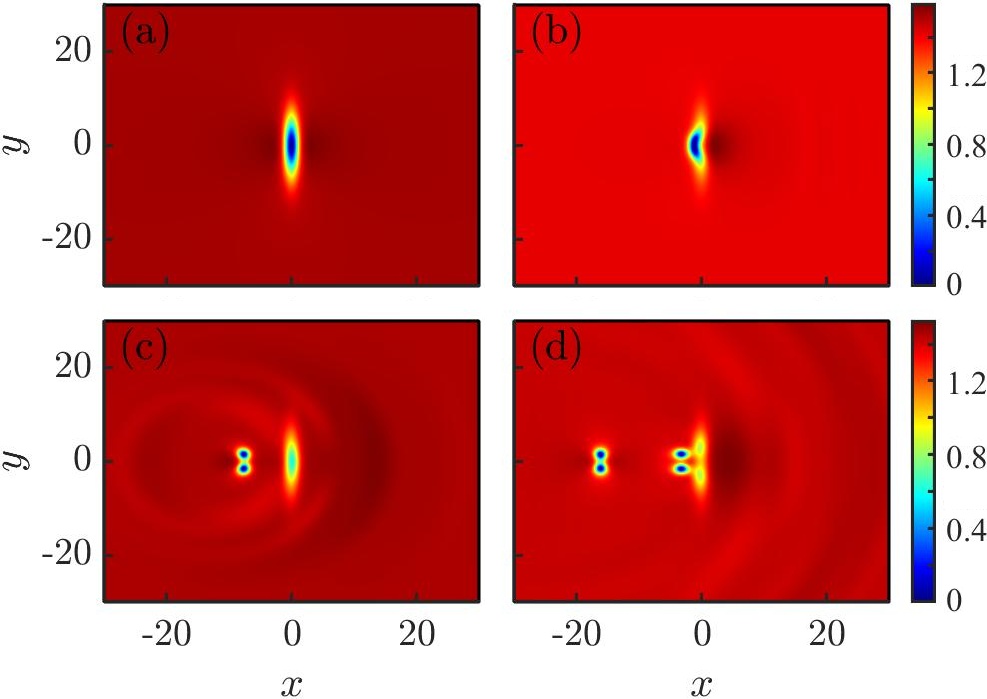}
\caption{
(Color online)
Evolution corresponding to a supercritical 2D defect running at 
velocity $c = 0.8$ for $\mu=0.5$ and $\lambda = 0.92$.
The defect impurity periodically emits a vortex-anti-vortex pair
in its wake (only two pairs shown here for $t\leq 55$).
(a) $t = 0$, (b) $t = 20$, (c) $t = 35$, and (d) $t = 55$.}
\label{fig:evol_2d}
\end{figure}

\begin{figure}[ht!]
\centering
\includegraphics[width=0.9\columnwidth]{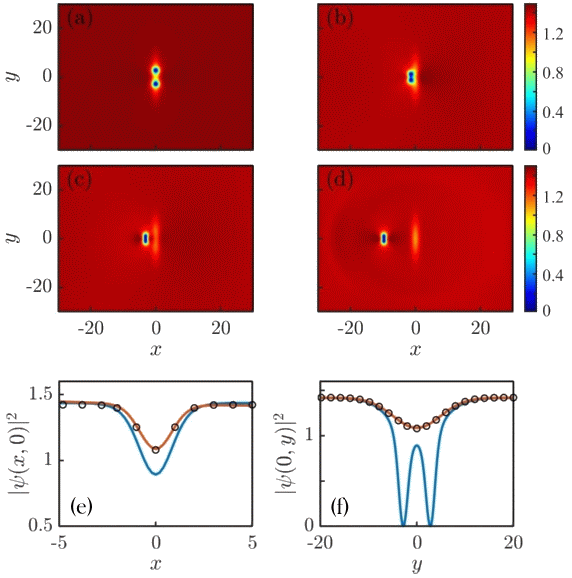}
\caption{
Evolution of the 2D unstable solution for a defect velocity
$c = 0.2$ below the speed of sound for $\mu=0.5$ and $\lambda=0.92$.
Panels (a)--(d) show the evolution where the deeper unstable
steady state solution sheds a vortex-anti-vortex pair and settles
to the stable shallower solution. 
(a) $t = 0$, (b) $t = 55$, (c) $t = 60$, and (d) $t = 75$.
Panel (e) and (f) depict the $y=0$ and $x=0$ cuts of the density
confirming that the initial unstable steady state (blue curves)
decay (after the emission of the vortex pair) to a shallower solution 
(red curves) that precisely corresponds to the stable steady state 
solution (shown by circles).
}
\label{fig:evol_2d_upper_2}
\end{figure}

As in 1D, for sufficiently large, supercritical, velocities, the 2D defect will not support a stable stationary state and will, accordingly,
emit a periodic train in its wake. In the 2D case, the defect produces a
train of vortex-anti-vortex pairs. A typical example depicting multiple
vortex shedding instances is shown in Fig.~\ref{fig:evol_2d} (larger 
times produce more vortex-anti-vortex shedding; results not shown here).
Also, as in the 1D case, the 2D model also features two steady state solutions
for subcritical defect speeds: an unstable relatively deep one and a stable
relatively shallow one.
Figure~\ref{fig:evol_2d_upper_2} confirms, similar to what we saw for
the 1D case, that for subcritical velocities, the deeper solution is 
unstable and, as it destabilizes, it sheds a {\em single} vortex-anti-vortex 
pair and eventually settles to the shallower stable steady state.

\begin{figure}[ht!]
\centering
\includegraphics[width=0.8\columnwidth]{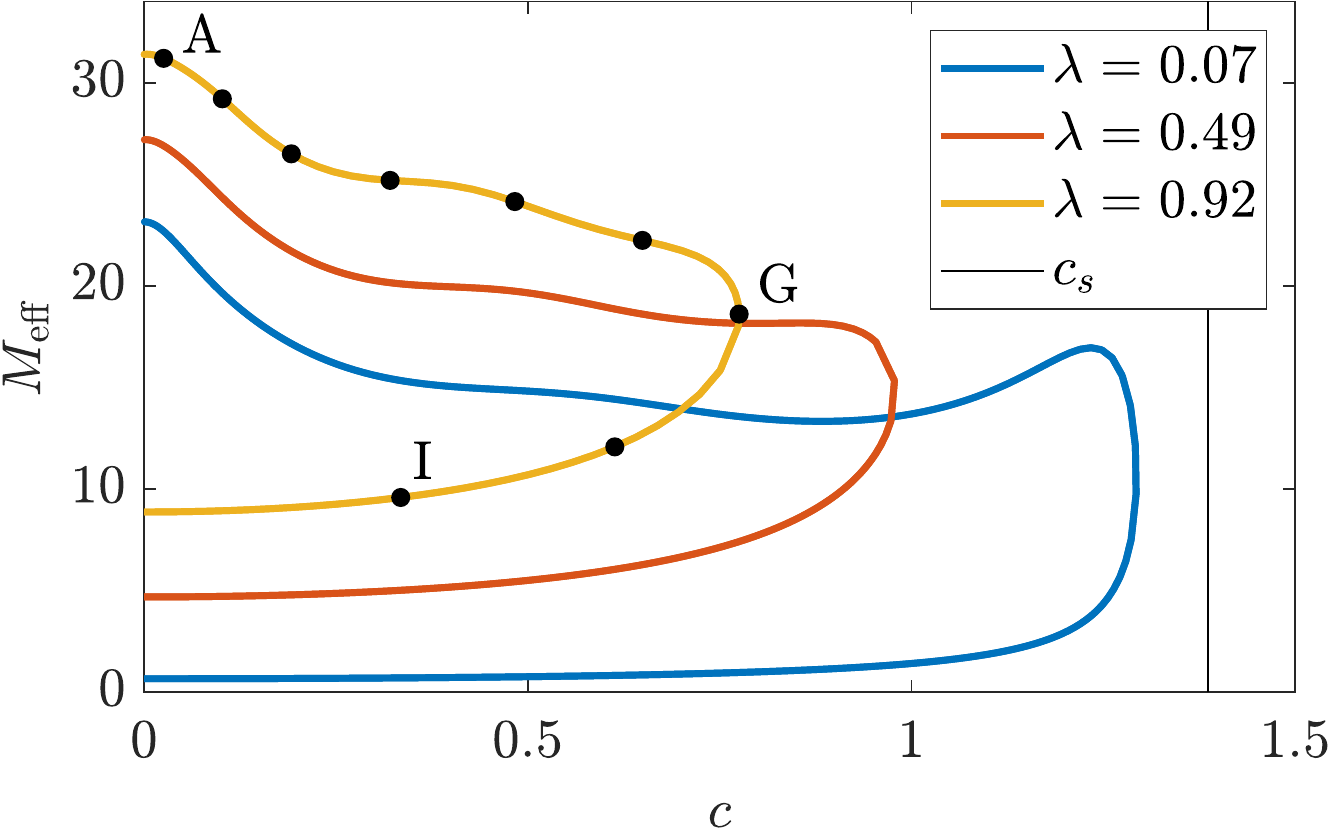}
\\[3.0ex]
\includegraphics[width=\columnwidth]{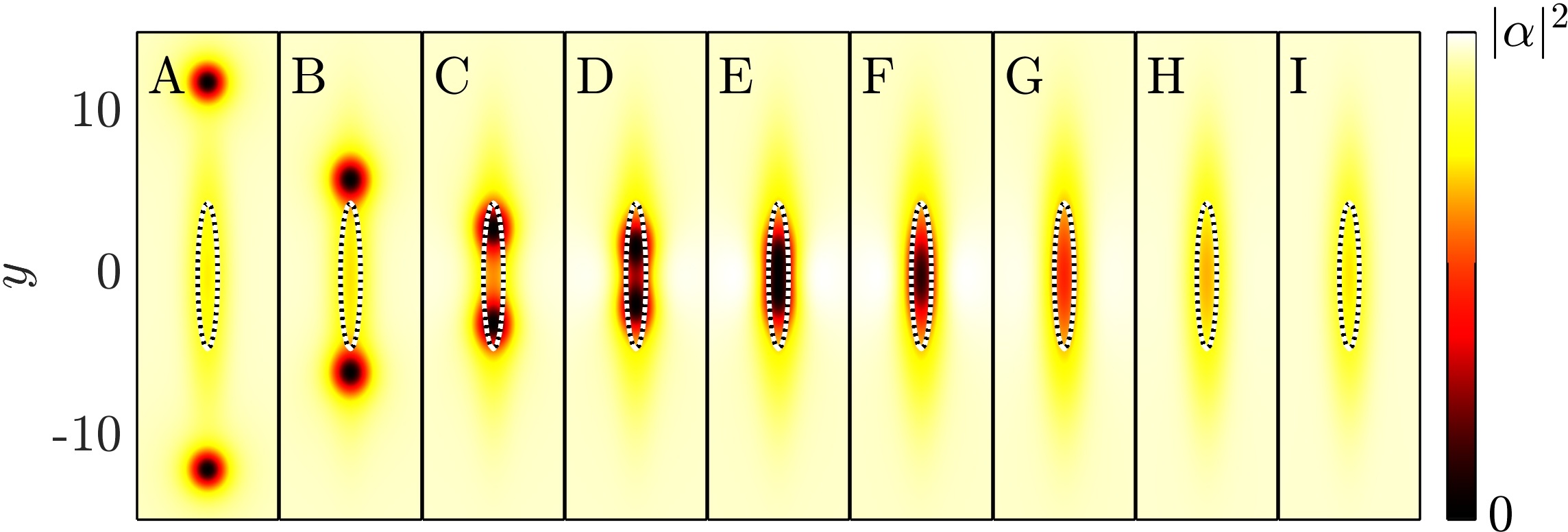}
\\[1.5ex]
\includegraphics[width=\columnwidth]{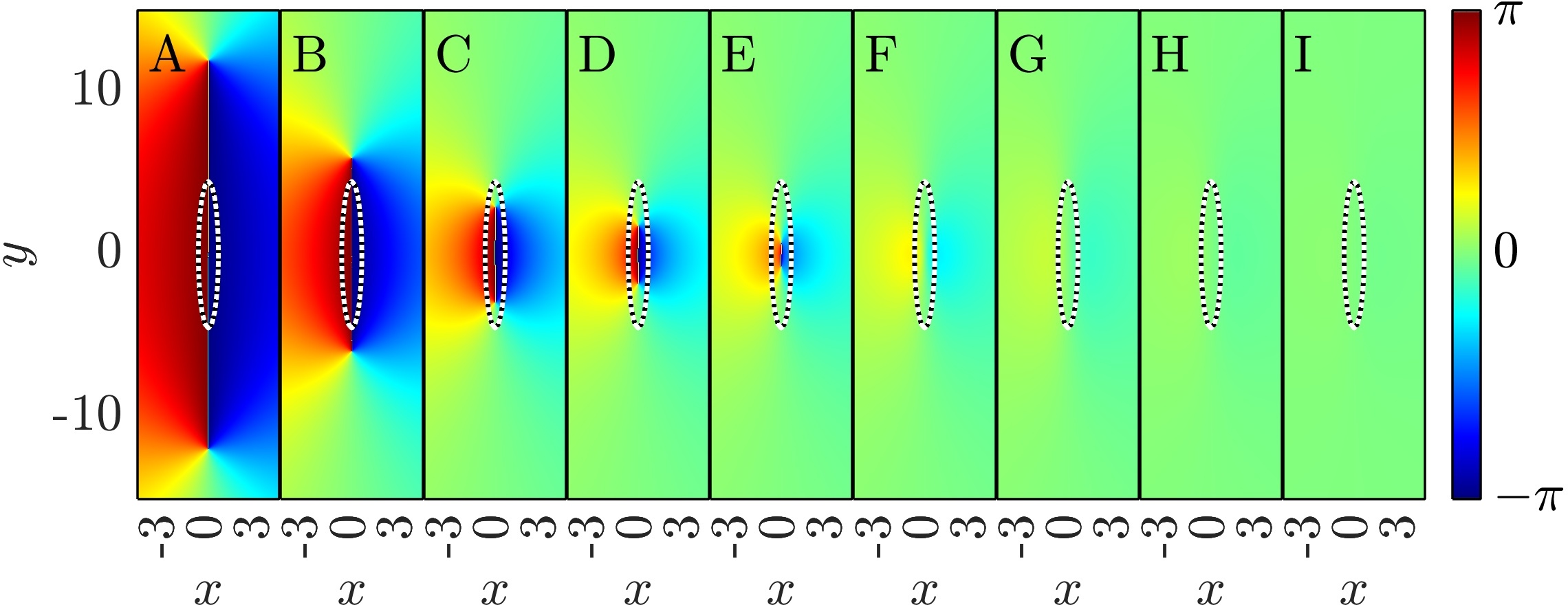}
\caption{%
(Color online)
Top:
Similar to Fig.~\ref{fig:bif_1d} but for the 2D case for $\mu=0.5$.
The vertical black line corresponds to the theoretical prediction of
speed of sound in Eq.~(\ref{eq:cs2D}).
Middle (bottom):
Corresponding density (phase) profiles at the different values of 
$c$ indicated by the black dots in the top panel. 
The dashed curve represents the isocontour level of the potential
$V_{\rm 2D}$ at 2/3 of its maximum height.
}
\label{fig:bif_2d}
\end{figure}

By monitoring the effective mass as in Eq.~(\ref{eq:Mass}), but replacing the
single $x$-integral by an integral over the whole 2D domain and adjusting the 
background as per Eq.~(\ref{eq:A2D}), we follow, using pseudo-arclength
continuation, the bifurcation diagram of subcritical 2D solutions.
The resulting bifurcation diagram for $\mu=0.5$ and for three different values 
of $\lambda$ is depicted in Fig.~\ref{fig:bif_2d}.
%
%
Interestingly, in this 2D case, the upper branch of steady state solutions 
contains, for small enough values of the defect speed (namely $c<0.52$), 
a vortex-anti-vortex pair (see panels A--E). This is precisely the 
vortical structure pair that detaches from this unstable solution as it is evolved
in time and finally settles to the corresponding stable lower branch solution,
which in turn lacks any vortices (see for instance the dynamical evolution
depicted in Fig.~\ref{fig:evol_2d_upper_2}).
It is also noteworthy that for small values of the running speed $c$,
the steady state vortex pair contains vortices that are relatively 
far away from the defect impurity (see for instance profile in panel 
A which corresponds to a velocity $c=0.025$). 
As $c$ increases, the vortices of the upper branch solution get closer to 
the defect impurity. Further increasing $c$ induces the vortices to get closer
to each other within the impurity until they merge and disappear
(in the present case of $\lambda=0.92$ at an approximate value of $c=0.52$).
Continuing past this vortex-merging point along the branch, the upper and
lower solutions bifurcate from each other (or, equivalently, terminate) in a saddle-center bifurcation
(case G in Fig.~\ref{fig:bif_2d}).

\begin{figure}[ht!]
\centering
\includegraphics[width=\columnwidth]{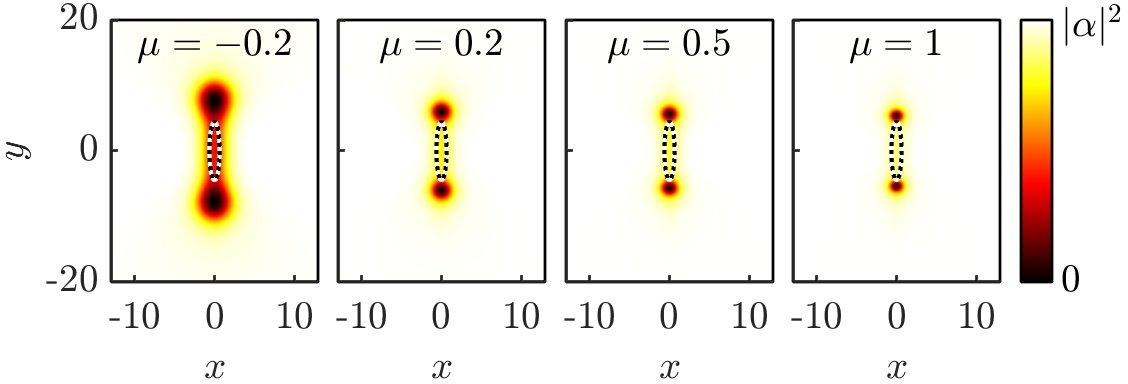}
\caption{Density profile of unstable solutions in the 2D case 
for $\lambda = 0.71$ and $c = 0.1$ for the different values of $\mu$
as indicated in the panels.
The dashed curve represents the isocontour level of the potential
$V_{\rm 2D}$ at 2/3 of its maximum height.
}
\label{fig:sols_2d_diffmu}
\end{figure}

The astute reader may have noticed the oscillatory behavior of
the effective mass for the upper branch solutions for small values
of $c$ in Fig.~\ref{fig:bif_2d}.
This oscillatory behavior is missing in the 1D case (see bifurcation
curves in Fig.~\ref{fig:bif_1d}). We attribute these oscillations to
the existence of the vortex pairs attached to the corresponding
steady states. As the vortices move closer to the impurity and run 
through it when $c$ is varied, they affect the effective mass and,
thus produce, these small oscillations.
In Fig.~\ref{fig:sols_2d_diffmu} we show the effects of varying 
the chemical potential $\mu$ on the upper branch steady state containing 
a pair of vortices attached at the end of the defect impurity.
The existence of vortices for a wide range of $\mu$ 
values suggests that the above-mentioned oscillations of the effective
mass will be visible for other values of the chemical potential.
Additionally, the phenomenology discussed herein is present for
different values of $\mu$. However, the main feature that chiefly
appears to change is the size of the vortices which shrinks, as the
healing length shrinks for increasing $\mu$. Moreover, for $\mu < 0$,
we can observe an implication of 
the droplet nature of the configuration and the effective
surface tension in such a setting, namely
a density modulation of the structure between the vortices,
a feature far less noticeable in the cases of larger $\mu$.

\begin{figure}[ht!]
\centering
\includegraphics[width=0.9\columnwidth]{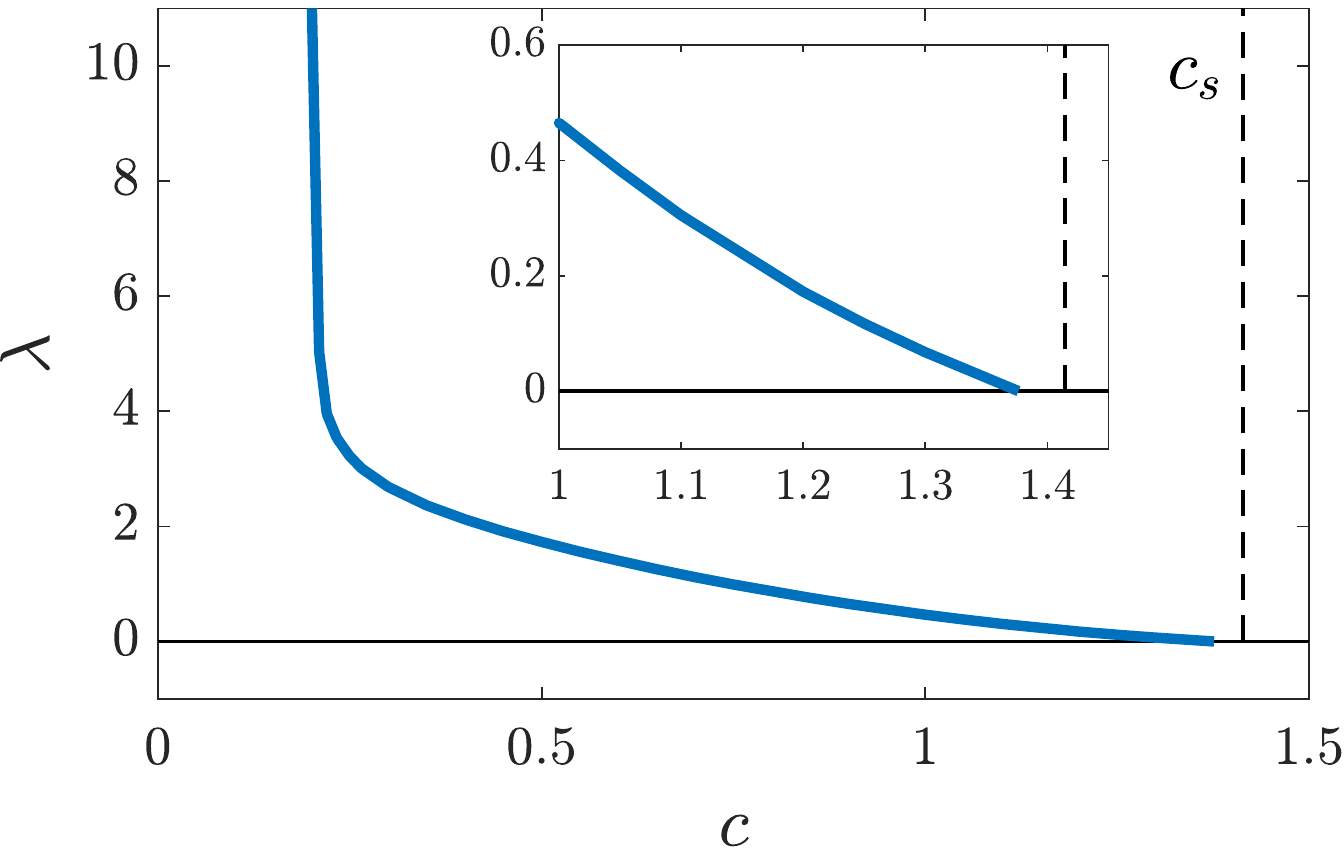}
\caption{
Critical values of defect strength $\lambda$ vs.~defect speed $c$
for the 2D case. Same layout as in Fig.~\ref{fig:lamc_1d}.
The vertical dashed line corresponds to the theoretical prediction of the
speed of sound in Eq.~(\ref{eq:A2D}), which is tantamount to the limit
of $\lambda \rightarrow 0$.
The inset shows a zoomed-in version for small defect strengths.
}
\label{fig:lamc_2d}
\end{figure}

Finally, let us now compare the theoretical estimation of the speed of sound 
as per Eq.~(\ref{eq:cs2D}) and the critical value of the velocity 
in 2D. 
For this purpose, we depict in Fig.~\ref{fig:lamc_2d}
the values of the defect strength $\lambda$ where the saddle-center
collision between the deep and shallow steady states occurs.
As the inset corroborates, the critical velocity 
coincides reasonably well with the theoretical speed of 
sounds $c_s$ when the defect strength $\lambda$ tends to zero, with the
difference being attributable to the approximations within the
numerical computations.

%
%
%
%
%

\subsection{Three-dimensional setting}
\label{sec:num:3D}

Let us now study the 3D model of Eq.~(\ref{eq:12}). In this case we
chose a defect impurity with the shape of a rectangular plate that is
thin in the direction of the defect movement (namely the $x$-direction) 
and relatively wide in the other two transverse directions.
Specifically, we use  the 3D defect 
{ $$
V_{\rm 3D}(x,y,z)=
V_{\rm 1D}(x)
\, H_{w_z}(z)\, H_{w_y}(y)
$$ 
with $\epsilon_x=1/4$, $w_y=8$ and $w_z=4$ and where $H_w$ is a
smoothed 1D top-hat function given by
$$
H_w(r)=\frac{1}{4}\left[\tanh\left(\frac{w}{2}+r\right)+1\right]
\left[\tanh\left(\frac{w}{2}-r\right)+1\right].
$$}
As it was the case for 1D and 2D setting above, we take a
relatively thin Gaussian profile along the direction of motion.
In the transverse directions we take a relatively large (when compared
to the thin longitudinal direction) rectangular plate.
One of the motivations to use a rectangular plate is to observe
the effects of this anisotropy in the transverse directions.
Naturally, an isotropic, namely circular, defect plate would give
rise to a perfectly isotropic steady state solution, which in turn, when
supersonic, would nucleate a symmetric vortex ring (results not shown here).

\begin{figure}[ht!]
\centering
\includegraphics[width=0.94\columnwidth]{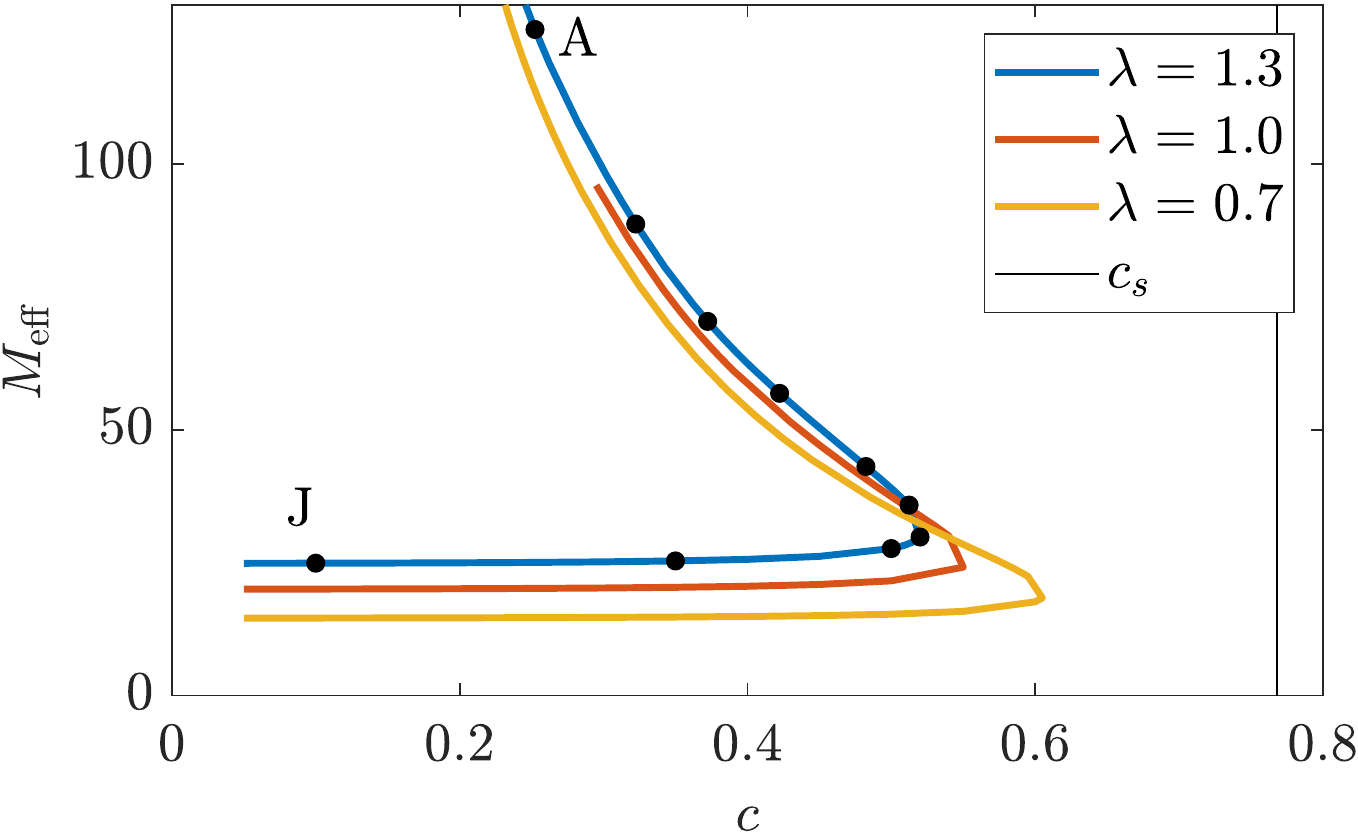}
\includegraphics[width=\columnwidth]{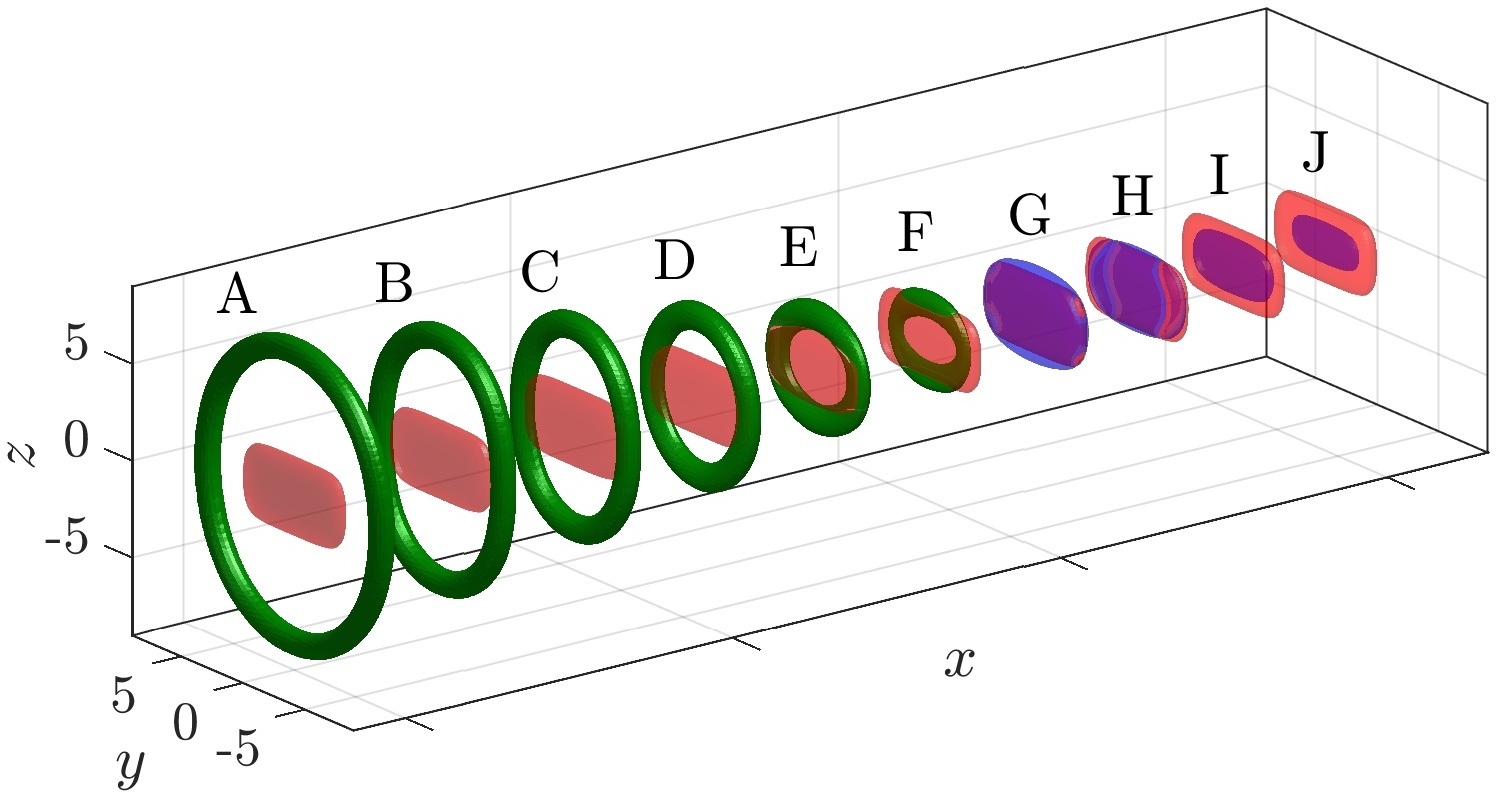}
\caption{
(Color online)
Top:
Bifurcation diagram for the stable and unstable branches in 3D for $g_1=1$ 
and $\mu=0.5$ and for the values of $\lambda$ indicated in the legend.
Similar layout and meaning as in Fig.~\ref{fig:bif_1d}.
The vertical black line corresponds to the theoretical prediction of
the 3D speed of sound in Eq.~(\ref{eq:cs3D}).
Bottom:
Corresponding profiles at the different values of $c$ indicated 
by the black dots in the top panel. Each instance depicts an isocontour
of the corresponding vorticity (solid green) together with 
an isocontour of the 3D defect potential $V_{\rm 3D}$ 
at 1/2 of its maximum amplitude (transparent red).
The points corresponding to G--J do not carry a sizeable
amount of vorticity and thus isocontours of the density are shown
instead (transparent blue).
}
\label{fig:bif_3d}
\end{figure}

\begin{figure}[ht!]
\centering
\includegraphics[height=3.4cm]{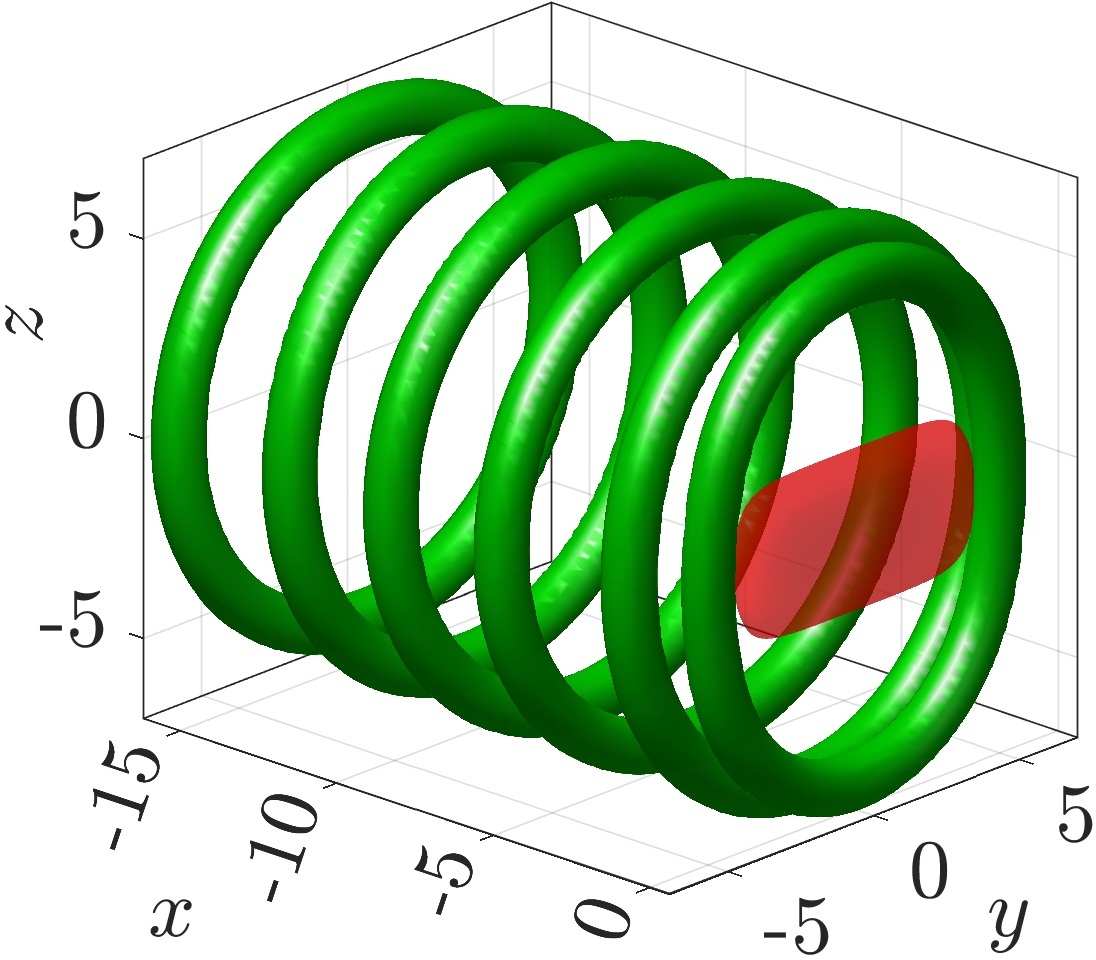}
~
\includegraphics[height=3.4cm]{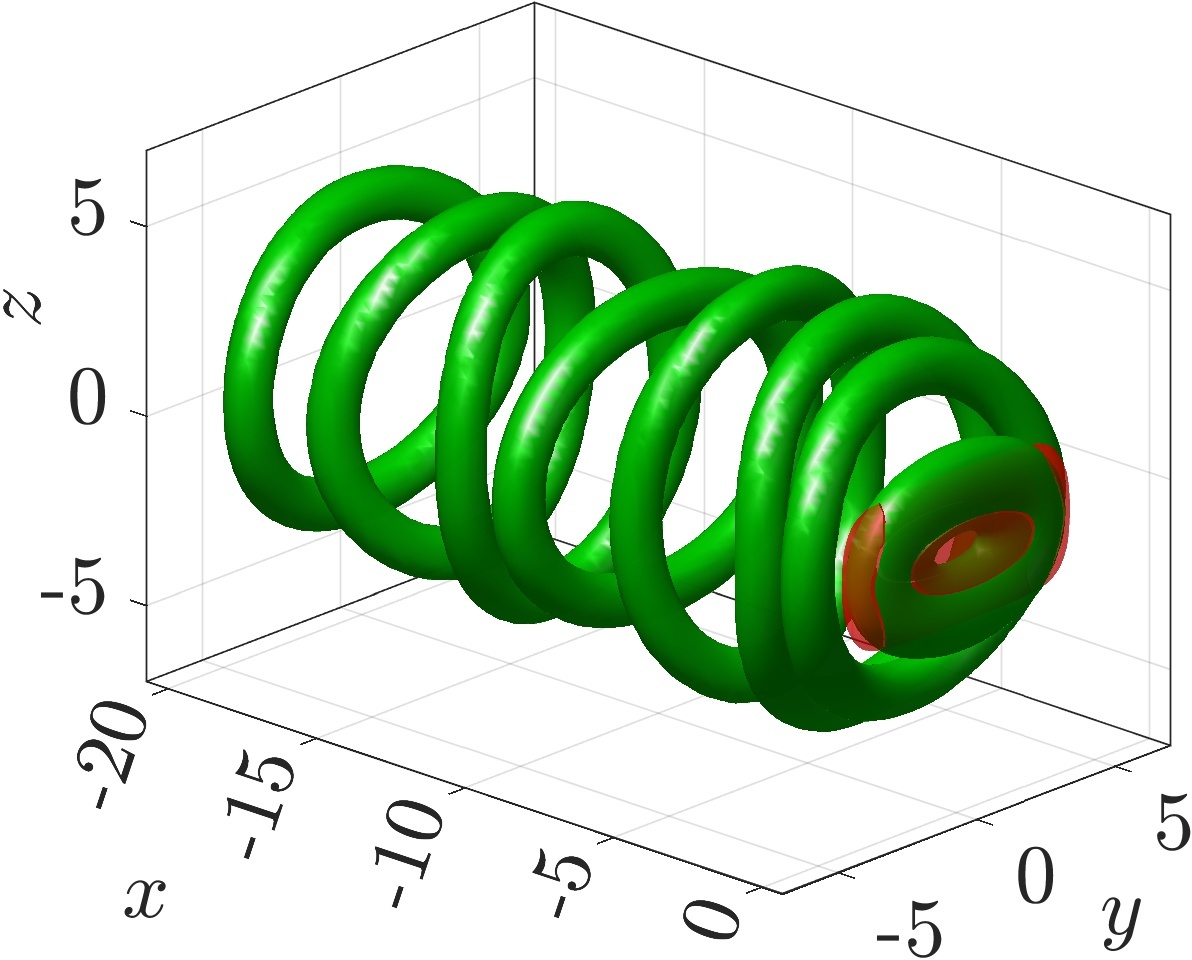}
\caption{
(Color online)
Examples of vortex ring emission from the unstable steady state solution. 
Each panel depicts overlaid snapshots of the vorticity (green surfaces) 
at different times during the emission of a single vortex ring from the 
defect (shown in red) on a co-moving reference frame.
The left and right panel corresponds, respectively, 
to the ensuing dynamics from cases B and F in Fig.~\ref{fig:bif_3d}.
Namely, for $\lambda=1.3$, $\mu=0.5$, $g_1=1$, and $c=0.3224$
(case B; left panel) and $c=0.5124$ (case F; right panel).
The times for the snapshots for cases B and F are, respectively,
$t=\{0,160,220,280,340,400\}$ and
$t=\{0,90,100,120,140,160,180,200\}$,
from the closest vortex ring to the defect to the one furthest away.
%
}
\label{fig:3D_emmission}
\end{figure}

Similarly to what we observe in the 1D and 2D cases, the 3D model
also gives rise to two branches of subcritical steady state solutions. 
These two solution branches correspond to unstable, relatively large
(when compared to the size of the defect), vortex rings
and stable density depletions.
We monitor the effective mass as in Eq.~(\ref{eq:Mass}), but replacing 
the single $x$-integral by an integral over the whole 3D domain and 
adjusting the background as per Eq.~(\ref{eq:A3D}).
As before, we use pseudo-arclength continuation to follow these
branches of subcritical 3D steady state solutions. The results 
for $g_1=1$ and {$\mu=0.5$} are depicted in Fig.~\ref{fig:bif_3d} for 
three different values of the defect strength $\lambda$.
It is interesting to notice that in the limit of small defect speed $c$, 
the upper branch of unstable solutions corresponds to large vortex
rings that are larger than, and seemingly detached from, the impurity.
On the other hand, the stable steady states, corresponding
to density depletions that do not carry vorticity, are tightly attached to
the defect. This provides a physical explanation for their corresponding
stability as the larger ring, being further away from the defect, 
detaches through the instability. On the other hand the smaller
density depletion is tightly attached to the impurity and is found 
to be dynamically stable.
Another interesting feature is that for intermediate
values of the velocity, the vortex ring has the opposite aspect ratio
than that of the defect plate. See for instance the vortex ring
corresponding to the points B and C which are taller than wider 
while the defect is, oppositely, wider than taller. 
Therefore, the effect of the anisotropy of the defect plate is to typically
create vortex rings that lack circular symmetry and that are slightly
``squeezed'' in the horizontal or vertical direction. This is to be
contrasted with the case of a circular plate that would nucleate a 
perfectly symmetric, circular, steady state and, in turn, shed a 
symmetric vortex ring (results not shown here).
The asymmetry present in the unstable steady state is inherited by the
vortex rings that are nucleated from the unstable subcritical steady state
as well as the train of vortex rings that are nucleated from the 
defect impurity running at supercritical speeds. This asymmetry means 
that the vortex rings nucleated by the defect will contain Kelvin
(vibrational) modes~\cite{KelvinModes1,KelvinModes2} that will induce 
internal oscillations (see Fig.~\ref{fig:3D_emmission}).

Let us now follow in more detail the nucleation of vortex rings
from the unstable branch of solutions. In particular, we depict in 
Fig.~\ref{fig:3D_emmission} the evolution of the unstable steady
states corresponding to cases B (left panel) and F (right panel) 
of Fig.~\ref{fig:bif_3d}. 
The simulations depict how the vortex ring, that is pinned by the 
defect, detaches (by virtue of the solution's instability) 
and then travels downstream. It is worth reminding that,
as per our setup in Sec.~\ref{sec:theory}, we are always mounted in 
a co-moving reference frame on top of the running defect.
As mentioned above, the pinned vortex rings are asymmetric as per
our choice of defect that has a rectangular transverse (i.e.,
perpendicular to its motion) cross section. Therefore, the detaching
vortex rings are not circular and thus are prone to Kelvin, internal,
oscillatory modes.
It is also interesting to note that the larger vortex rings,
corresponding to relatively small values of the running defect 
speed $c$, detach and keep their relatively large radius
---cf.~case B corresponding to the left panel in 
Fig.~\ref{fig:3D_emmission}.
On the other hand, for larger defect speeds, the pinned vortex 
ring has a small radius and thus expands as it detaches
---cf.~case F corresponding to the right panel in 
Fig.~\ref{fig:3D_emmission}.
It is also worth mentioning that the detached vortex rings do
not travel downstream at velocity $c$ with the background
fluid velocity. This is because the detached vortices (and also 
the nucleated vortices for supercritical speeds) have an
intrinsic velocity $v_i$ that goes {\em against} the background flow
(see Ref.~\cite{CaplanVR} and references therein).
Thus, vortex rings do travel downstream but with a slower
velocity than the background flow corresponding to $c-v_i$.

\begin{figure}[ht!]
\centering
\includegraphics[width=0.94\columnwidth]{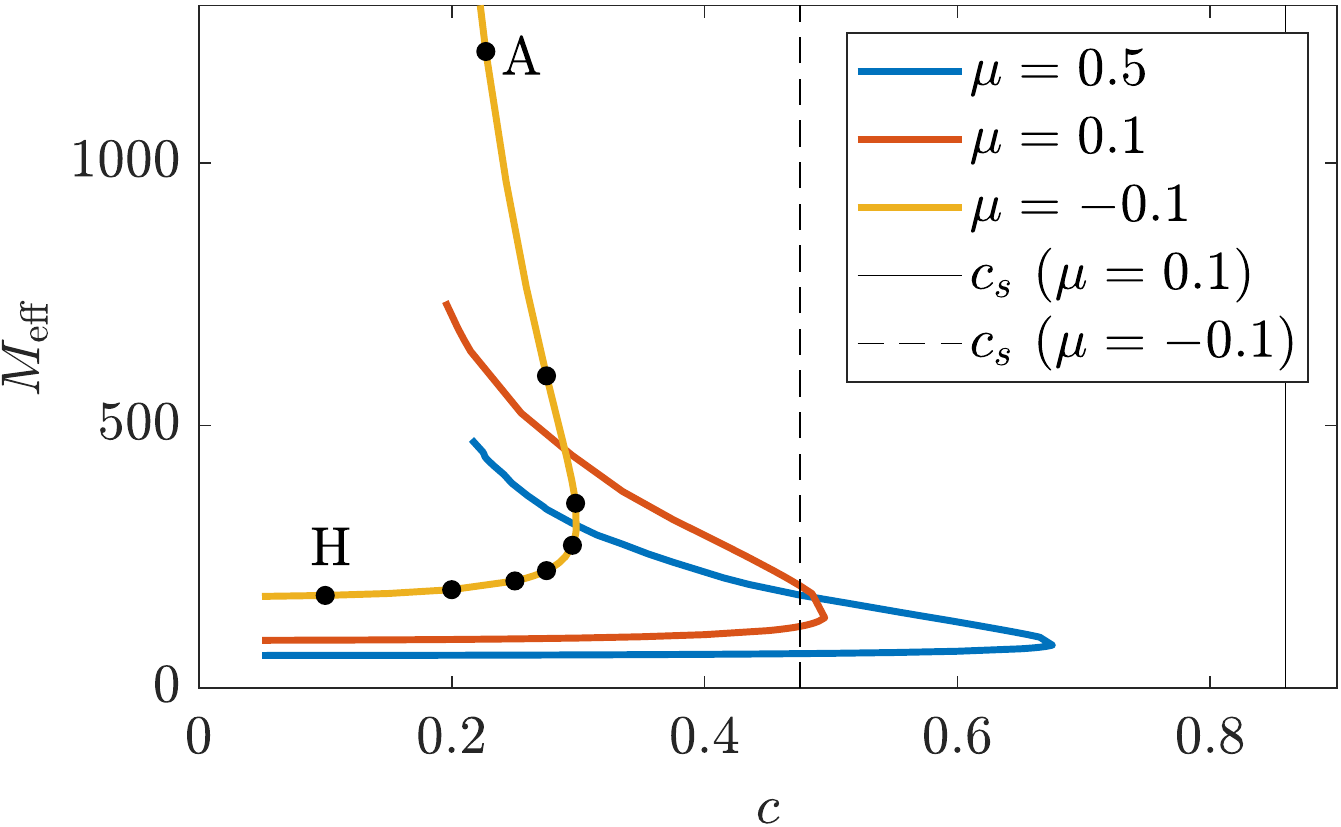}
\includegraphics[width=\columnwidth]{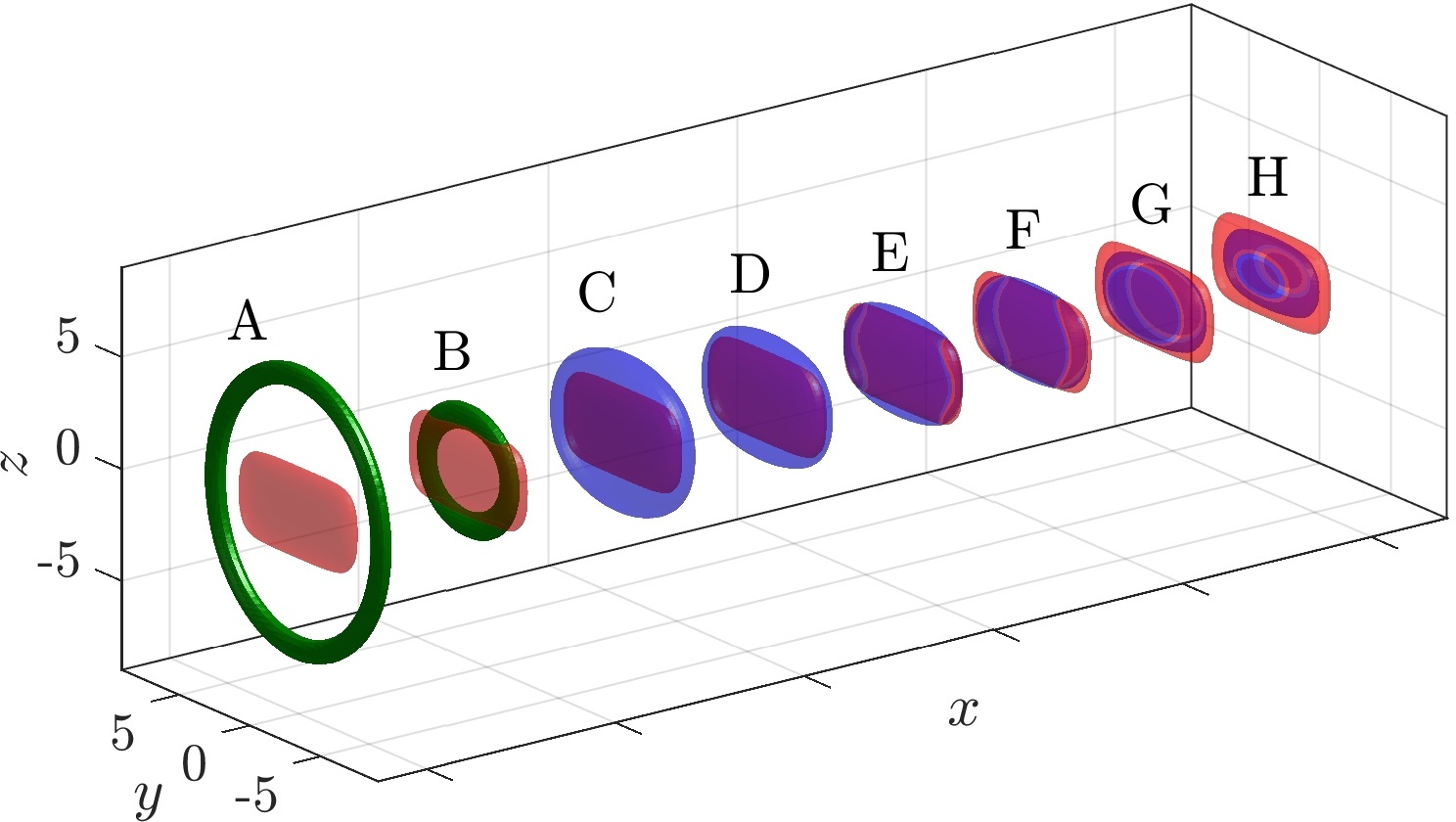}
\caption{
(Color online)
Same as in Fig.~\ref{fig:bif_3d} but for constant $\lambda=1.3$ and $g_1=-1$
for the values of $\mu$ indicated in the legend.
In this case, the points corresponding to C--H do not carry a sizeable
amount of vorticity and thus density isocontours are shown instead.}
\label{fig:mubif_3d}
\end{figure}

Finally, since it may be physically relevant to consider negative values
of $g_1$ in the 3D model~(\ref{eq:12}), we depict in Fig.~\ref{fig:mubif_3d}
the bifurcation diagram for the steady state solutions for $g_1=-1$ and
a defect strength of $\lambda=1.3$ for three different values of the
chemical potential $\mu$.
The results are similar as the ones presented for $g_1=1$ in Fig.~\ref{fig:bif_3d}
with no significant qualitative differences.

\section{Conclusions \& Future Challenges}
\label{sec:conclu}

We have studied the effects of running an impurity defect through a
Bose-Einstein condensate in a regime that takes into account the
incorporation of the Lee-Huang-Yang (LHY) correction that gives
rise to quantum droplets.
We systematically explored the 1D, 2D, and 3D cases that feature
different nonlinear LHY corrections.
For all dimensionalities, we followed, using pseudo-arclength continuation,
the two subcritical solution branches that exist for defect velocities 
below the critical speed and which feature a localized waveform
co-traveling with the defect.
%
These solutions correspond to dark solitons, vortex-anti-vortex pairs,
and vortex rings for, respectively, the 1D, 2D, and 3D cases.
In all cases we find that there exist an upper and lower branch of
solutions connected through a saddle-center bifurcation. The upper
branch is found to always be unstable and corresponds to a solution
that has a relatively larger effective mass and a nonlinear
state further detached from the defect. In contrast, the
lower solution branch of relatively smaller effective mass is dynamically
stable for subcritical velocities and features a state more
closely bound to the defect.
When dynamically evolved, the solution on the upper branch always
destabilizes by shedding a single coherent structure and then settling
to its lower branch, stable, sibling solution.
By using a perturbation approach we are able to theoretically predict the
corresponding speed of sound of the medium which, in turn, takes a different
functional form for the different model dimensionalities.
We corroborate that the theoretically computed speed of sounds does
match the bifurcation point (where the upper and lower branches collide)
as the strength of the defect tends to zero and we identify
how this critical speed (and the corresponding
saddle-center bifurcation point) deviates, i.e., decreases from this threshold, as we move into the finite defect strength case.
In the 3D case we showcase the effect of using an anisotropic defect
in the transverse direction. This anisotropy is responsible for the
formation of ``squeezed'' vortex rings again featuring a similar
bifurcation diagram. 
We explored the relevant phenomenologies not only as a function 
of the dragging speed and the defect strength, but also in terms
of the chemical potential variations, observing the variation
of the states as the droplet (negative chemical potential) limit
is approached. Finally, we also performed dynamics in the 
supercritical case, observing how the disappearance of the
states co-traveling with the defect results in the emission of
a train of dark solitons, or a street of vortices or an array
of vortex rings in 1D, 2D and 3D, respectively.

The present work naturally suggests numerous additional
considerations in the context of nonlinear wave patterns
embedded in the types of droplet models that were considered
herein. While some studies along this vein have recently
taken place both in 1D~\cite{Kartashov_2022,edmonds} 
and in higher
dimensions~\cite{luo2020new}, it appears that a detailed
understanding of dark solitons, vortices or vortex pairs
and vortex rings embedded within droplet configurations is largely
still missing, as is a characterization of their stability
and dynamics. We believe that such a systematic study and
also comparison with beyond mean-field models (see for 
a recent review Ref.~\cite{mistakidis2022cold}) would be of
particular interest in future work. Such topics are
presently under consideration and will be reported in future
publications.

\appendix

\section{Speed of sound in 2D}
\label{app:2D}

Looking for solutions to Eq.~(\ref{eq:6}) of the form 
of Eq.~(\ref{eq:ARphi}) but in 2D yields
\begin{eqnarray}
cR_x &=& \frac{1}{2} (2R_x\phi_x+R\phi_{xx}+2R_y\phi_y+R\phi_{yy}),
\label{eq:7}
\\[1.0ex]
cR\phi_x&=&-\frac{1}{2} (R_{xx}+R_{yy}-R\phi_x^2-R\phi_y^2)
\notag
\\[1.0ex]
\label{eq:8}
&&+R^3\ln{R^2}-\mu R+VR.
\end{eqnarray}
Further linearizing and expanding the phase and density terms by using  
transverse modes of wavenumber $k$ in the $y$-direction as
$\phi(x,y,t)=\epsilon\, \theta(x,t)e^{iky}$ and 
$R(x,y,t)=|\alpha|+\epsilon\, r(x,t)e^{iky}$, 
Eqs.~(\ref{eq:7}) and (\ref{eq:8}) become 
\begin{eqnarray}
{2cr_x} &=& |\alpha|\left(\theta_{xx} - k^2\theta\right),
\label{eq:9}
\\[1.0ex]
r_{xx} &=& -2c|\alpha|\theta_x+ r(k^2+4|\alpha|^2+4\mu),
\label{eq:10}
\end{eqnarray}
Finally, by making the substitutions $r = ae^{\Lambda x}$ and 
$\theta = be^{\Lambda x}$, Eqs.~(\ref{eq:7}) and (\ref{eq:8}) read
$$
A
\begin{bmatrix}
a\\
b
\end{bmatrix}
= 
\begin{bmatrix}
0\\
0
\end{bmatrix}
,
$$
where
\[A = \begin{bmatrix}
{2c\Lambda}/{|\alpha|} &~~ k^2-\Lambda^2\\[2.0ex]
k^2+4|\alpha|^2+4\mu-\Lambda^2 &~~ -2|\alpha|\Lambda.
\end{bmatrix}\]
Then, setting $\det(A)= 0$ with $k = 0$ and $\Lambda = 0$ 
yields the speed of sound for the 2D setting as
\begin{equation}
c_s = \sqrt{|\alpha|^2+\mu}.
\label{eq:app:11}
\end{equation}
We test this prediction numerically under different settings
in Sec.~\ref{sec:num:2D}.

\section{Speed of sound in 3D}
\label{app:3D}

Looking for solutions to Eq.~(\ref{eq:6}) of the form
of Eq.~(\ref{eq:ARphi}) but now in 3D yields
\begin{eqnarray}
2cR_x &=& 2R_x\phi_x + R\phi_{xx} + 2R_y\phi_y + R\phi_{yy} 
\notag
\\[1.0ex]
\label{eq:13}
&&+ 2R_z\phi_z + R\phi_{zz}
\\[1.0ex]
cR\phi_x &=& -\frac{1}{2} (R_{xx} - R\phi_x^2 + R_{yy} - R\phi_y^2 + R_{zz} - R\phi_z^2) 
\notag
\\
\label{eq:14}
&&+ R^4 + g_1 R^3 - \mu R + VR
\end{eqnarray}
As in the 2D calculations, we now allow for the density and phase
terms to contain transverse modes. In this case, including modes 
in the $y$ and $z$ directions as
$\phi(x,y,z,t)=\epsilon\, \theta(x,t)e^{ik_yy}e^{ik_zz}$ and 
$R(x,y,t)=|\alpha|+\epsilon\, r(x,t)e^{ik_yy}e^{ik_zz}$.
Then, after linearization, Eqs.~(\ref{eq:13}) and (\ref{eq:14}) become
\begin{eqnarray}
{2cr_x} &=&{|\alpha|}\left[ \theta_{xx} - (k_y^2+k_z^2)\theta\right],
\label{eq:15}
\\[1.0ex]
r_{xx} &=& r\left[2|\alpha|^3+4\mu+(k_y^2+k_z^2)\right]-2c|\alpha|\theta_x.
\label{eq:16}
\end{eqnarray}
Using the substitutions $r = ae^{\Lambda x}$ and $\theta = be^{\Lambda x}$
yields the linear system
$$
A
\begin{bmatrix}
a \\
b
\end{bmatrix} = 
\begin{bmatrix}
0 \\
0
\end{bmatrix},
$$
where
\[A = \begin{bmatrix}
{2c\Lambda}/{|\alpha|} &~~ (k_y^2+k_z^2)-\Lambda^2\\[2.0ex]
(k_y^2+k_z^2)+2|\alpha|^3+4\mu-\Lambda^2 &~~ -2c|\alpha|\Lambda
\end{bmatrix}\]
Finally, setting $\det(A)= 0$ and $k_y=k_z=0$ and $\Lambda = 0$
yields the expression for the speed of sound in the 3D setting
\begin{equation}
c_s = \sqrt{\frac{|\alpha|^3+2\mu}{2}}.
\label{eq:app:17}
\end{equation}

\section{Speed of sound for the NLS}
\label{app:NLS}

For comparison with the LHY case and for completeness, the next two
sections show the results of the pure NLS model case:
\begin{equation}
\label{eq:NLS}
    i\partial_t A -ic\,\partial_xA= -\frac{1}{2}\nabla^2 A + |A|^2A - \mu A + V\,A,
\end{equation}
with $\nabla^2$ being the Laplacian and $V$ the corresponding defect potential.
In contrast with the case with the LHY correction, the pure NLS model
admits a single homogeneous steady state of density $|\alpha|^2=\mu$.

\subsection{NLS case in 1D}
\label{app:1D_NLS}

Analyzing this case in a way similar to the one in Sec.~\ref{sec:theory:1D}, indeed following the 
work of Ref.~\cite{hakim},
after replacing $A(x) = R(x)e^{i\phi(x)}$, the system
\begin{eqnarray}
\phi_xR^2 &=& \frac{1}{2}cR^2+C_I,
\label{eq:app1a}
\\[1.0ex]
\label{eq:app1b}
cR\phi_x&=&- (R_{xx}-R\phi_x^2)+R^3-\mu R+VR,
\end{eqnarray}
with the constant of integration $C_I=-\frac{1}{2}c\mu$.
Solving for $\phi_x$ in Eq.~(\ref{eq:app1a}) and replacing in it in
Eq.~(\ref{eq:app1b}) yields
$$
R_{xx} = \frac{1}{4}c^2\left(-R+\frac{\mu^2}{R^3}\right) +R^3-\mu R+VR.
$$
Linearizing as before, using $R(x) = |\alpha| + r(x)$, yields the following
expression for the evolution of the perturbation:
$$
  r_{xx}=r(2\mu-c^2),
$$
and, therefore, the speed of sound in 1D for the pure NLS case is given by
\begin{equation}
\label{eq:cs1D_NLS}
c_s=\sqrt{2\mu}.
\end{equation}
The relevant calculation both in this 
section and in the next one is provided for
reasons of completeness.

\begin{figure}[ht!]
\centering
\includegraphics[width=0.8\columnwidth]{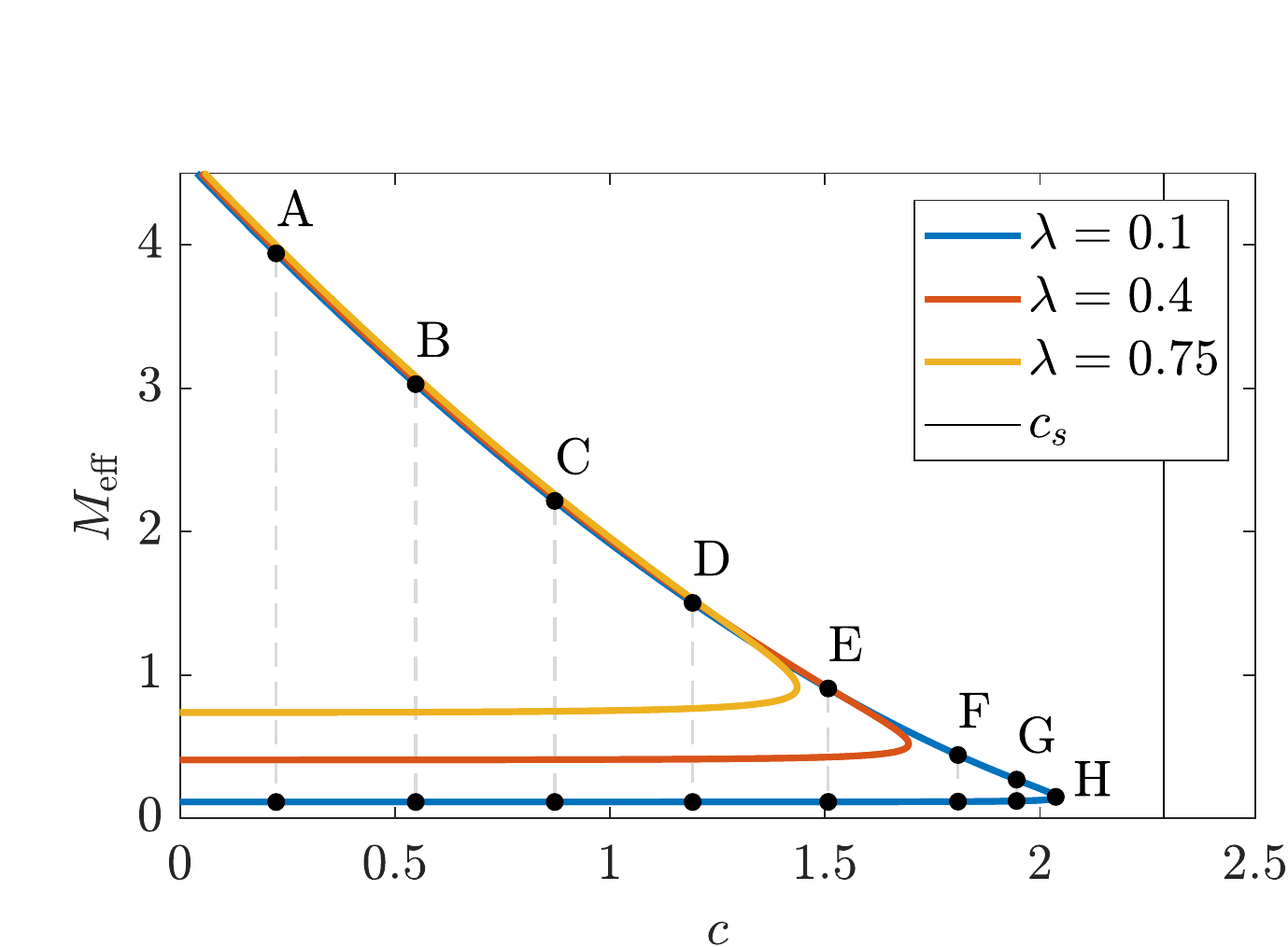}
\\[3.0ex]
\includegraphics[width=\columnwidth]{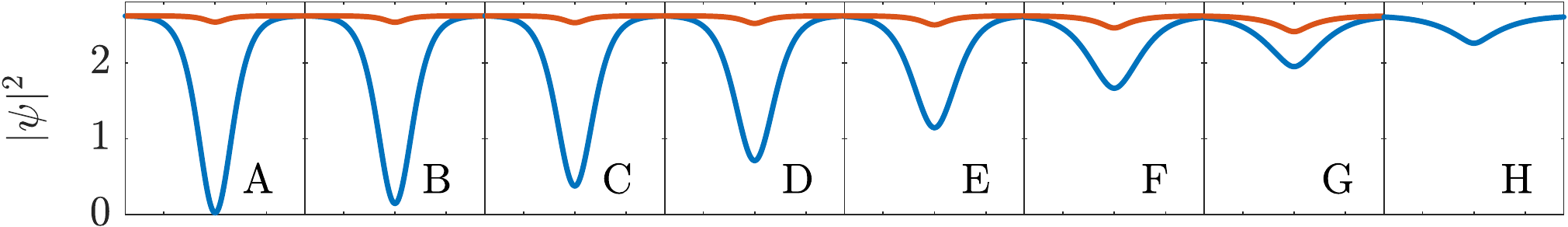}
\\[2.0ex]
\includegraphics[width=\columnwidth]{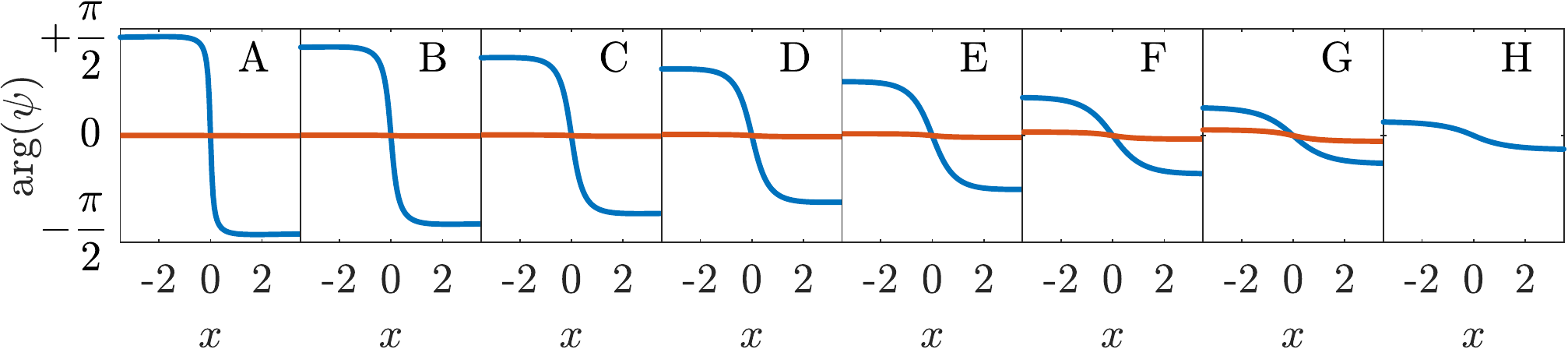}
\caption{(Color online)
Same as in Fig.~\ref{fig:bif_1d} but for the standard
(cubic) NLS model (\ref{eq:NLS}) 
in 1D without the LHY correction.
The value of $\mu=|\alpha|^2$ was chosen here so as to match the background
density supporting the dark solitons for the LHY case.
In this case, the theoretical prediction for the speed of sound is
given by Eq.~(\ref{eq:cs1D_NLS}).
Note the similarity with the results of the LHY case in Fig.~\ref{fig:bif_1d},
albeit with a different scale in the $c$-axis.
}
\label{fig:bif_1d_NLS}
\end{figure}

In Fig.~\ref{fig:bif_1d_NLS} depicts the numerical results corresponding
to the standard 1D NLS model (\ref{eq:NLS}). These results are to be compared
with the corresponding ones in Fig.~\ref{fig:bif_1d} that include the LHY 
correction as per Eq.~(\ref{eq:1}).
When comparing the two  cases at first glance, little difference is observed.
However, do note the quite different scales of defect speeds $c$ such that
the velocities for the standard NLS case are about twice of those with the
LHY correction.

\begin{figure}[ht!]
\centering
\includegraphics[width=0.8\columnwidth]{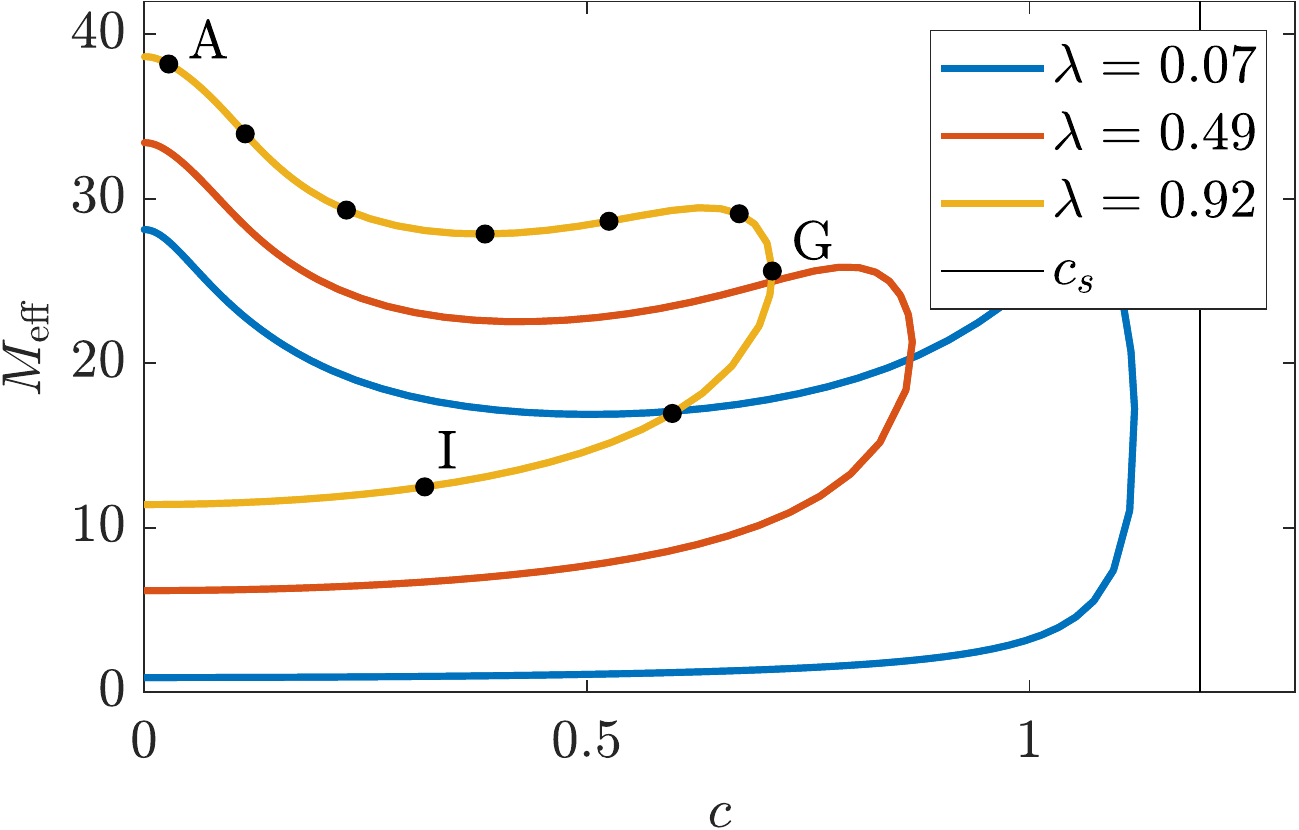}
\\[3.0ex]
\includegraphics[width=\columnwidth]{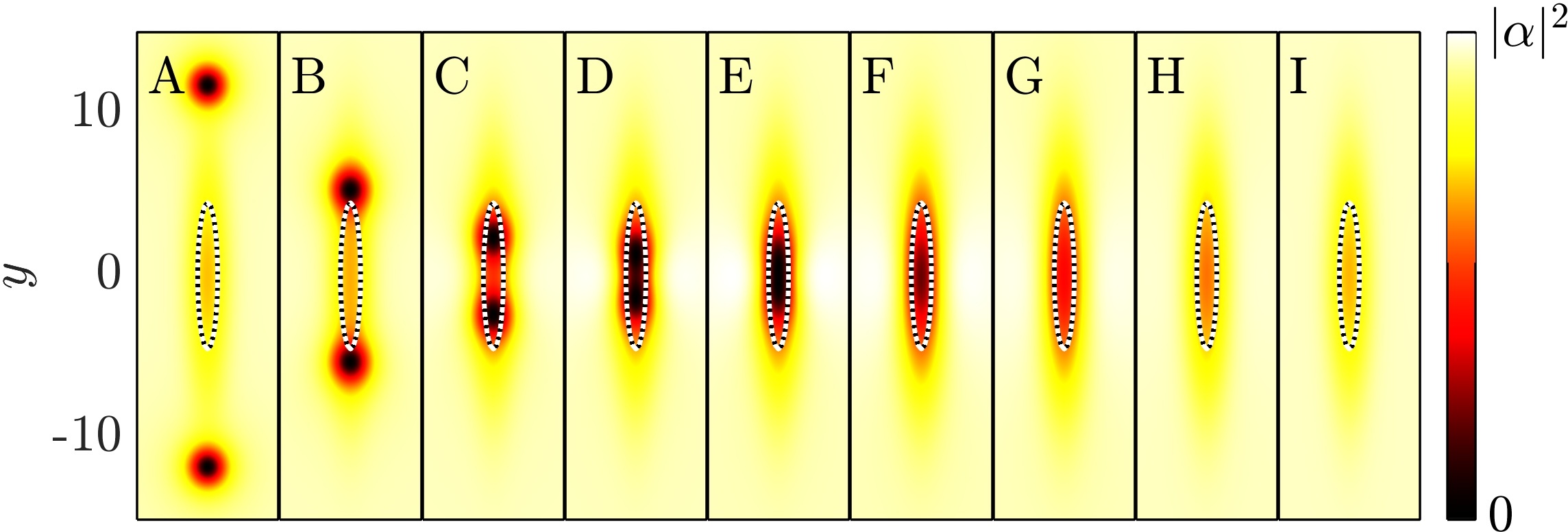}
\\[1.5ex]
\includegraphics[width=\columnwidth]{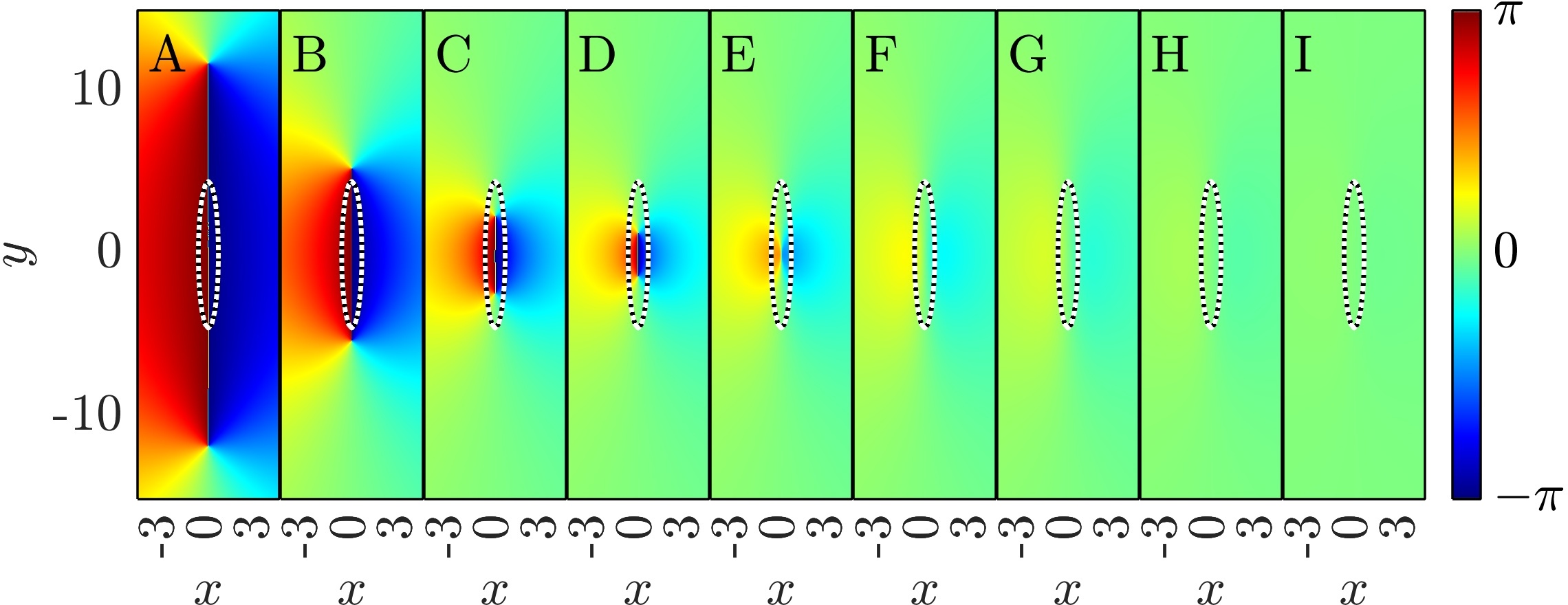}
\caption{%
(Color online)
Similar to Fig.~\ref{fig:bif_2d} but for the standard NLS 2D model
without the LHY correction.
The vertical line in the top panel is the predicted speed of sound
for the pure NLS in 2D given by Eq.~(\ref{eq:cs2D_NLS}).
}
\label{fig:bif_2d_NLS}
\end{figure}

\subsection{NLS case in 2D}
\label{app:2D_NLS}

Following a very similar analysis to the one in Sec.~\ref{sec:theory:2D} and 
Appendix~\ref{app:2D} yields, after (i) separating density and phase by 
$A(x,y,t) = R(x,y,t)e^{i\phi(x,y,t)}$, (ii) expanding in transversal modes
by using $\phi(x,y,t)=\epsilon \theta(x,t)e^{iky}a$, (iii) linearizing
the perturbed solution using $R(x,y,t)=r_0+\epsilon r(x,t)e^{iky}$,
and (iv) some algebra, yields, to first order the system:
\begin{eqnarray}
\frac{2cr_x}{r_0} &=& \theta_{xx} - k^2\theta,
\\[1.0ex]
r_{xx} &=& -2cr_0\theta_x+ r(k^2+4\mu).
\end{eqnarray}
Then, making the substitutions $r = ae^{\lambda x}$ and 
$\theta = be^{\lambda x}$, yields the matrix system
%
%
which induces the following matrix system
$$
\begin{bmatrix}
\frac{2c\lambda}{r_0} & k^2-\lambda^2
\\[1.ex]
k^2+4\mu-\lambda^2\quad & -2cr_0\lambda
\end{bmatrix}
\begin{bmatrix}
a\\[1.ex]
b
\end{bmatrix}
= 
\begin{bmatrix}
0\\[1.ex]
0
\end{bmatrix}.
$$
Setting the determinant equal to zero then yields the characteristic
polynomial
$$-\lambda^4 + \lambda^2(-4c^2+2k^2+4\mu)-k^2(k^2+4\mu)=0,$$
which implies that, for the most unstable mode $k=0$, the speed of
sound for the standard NLS model in 2D is given by
\begin{equation}
\label{eq:cs2D_NLS}
c_s=\sqrt{\mu}.
\end{equation}

Figure~\ref{fig:bif_2d_NLS} depicts the numerical results corresponding
to the standard 2D NLS model (\ref{eq:NLS}). These results are very 
similar to the corresponding ones in Fig.~\ref{fig:bif_2d} 
that include the LHY correction as per Eq.~(\ref{eq:6}).
As it was the case for the droplet model,
the NLS also displays a non-monotonic behavior of the 
effective mass as $c$ decreases for the upper branch.
We again attribute this behavior to the appearance of the
vortices that start, for small $c$, far away from the impurity and
then get closer to it as $c$ increases (see middle set of
panels in Fig.~\ref{fig:bif_2d_NLS}).

%


\providecommand{\noopsort}[1]{}\providecommand{\singleletter}[1]{#1}%

\end{document}